\documentclass[sn-mathphys,Numbered]{sn-jnl}

\usepackage{graphicx}%
\usepackage{multirow}%
\usepackage{amsmath,amssymb,amsfonts}%
\usepackage{amsthm}%
\usepackage{mathrsfs}%
\usepackage[title]{appendix}%
\usepackage{xcolor}%
\usepackage{textcomp}%
\usepackage{manyfoot}%
\usepackage{booktabs}%
\usepackage{algorithm}%
\usepackage{algorithmicx}%
\usepackage{algpseudocode}%
\usepackage{listings}%
\usepackage{tikz}
\usepackage{pgfplots}
\usepgfplotslibrary{groupplots}
\usepackage{siunitx}
\usepackage{booktabs}
\usepackage{subcaption}
\usepackage{threeparttable}
\usepackage{adjustbox}
\usepackage{hhline}
\usepackage{booktabs}
\usepackage{cancel}
\usepackage{xcolor}
\usepackage[
  colorinlistoftodos,
  backgroundcolor=green,
  linecolor=green,
  textsize=footnotesize
]{todonotes}

\theoremstyle{thmstyleone}%
\theoremstyle{thmstyletwo}%

\theoremstyle{thmstylethree}%

\definecolor{myjunker}{rgb}{0.82, 0.1 0.26}
\definecolor{mytobi}{rgb}{0.0, 0.0, 1.0}
\definecolor{mycem}{rgb}{0.0, 0.5, 0.0}
\newcommand{\bigcdot}{\mathbin{\text{\large$\bullet$}}}

\begin{document}

\title[Article Title]{An energy-based material model for the simulation of shape memory alloys under complex boundary value problems}


\author[]{ \fnm{Cem} \sur{Erdogan*} }\email{*erdogan@ikm.uni-hannover.de}
\author[]{\fnm{Tobias} \sur{Bode}}
\author[]{\fnm{Philipp} \sur{Junker}}
\affil{%
  \begin{minipage}{\textwidth}
    \hspace{30mm}
    \orgname{*Leibniz University Hannover},\orgdiv{ Institute of Continuum Mechanics} 
  \end{minipage}%
  \vspace{1pt}  
  \begin{minipage}{\textwidth}
   \hspace{30mm}
   \orgaddress{\street{An der Universität 1}, \city{Garbsen}, \postcode{30823}, \state{lower saxony}, \country{Germany}}
  \end{minipage}%
  \vspace{-7mm}
}




\abstract{Shape memory alloys are remarkable 'smart' materials used in a broad spectrum of applications, ranging from aerospace to robotics, thanks to their unique thermomechanical coupling capabilities. Given the complex properties of shape memory alloys, which are largely influenced by thermal and mechanical loads, as well as their loading history, predicting their behavior can be challenging. Consequently, there exists a pronounced demand for an efficient material model to simulate the behavior of these alloys. This paper introduces a material model rooted in Hamilton's principle. The key advantages of the presented material model encompass a more accurate depiction of the internal variable evolution and heightened robustness. As such, the proposed material model signifies an advancement in the realistic and efficient simulation of shape memory alloys.}




\maketitle
\section{Introduction}\label{sec1}
\noindent Shape memory alloys represent unique materials classified within the realm of 'smart materials', a subset that falls under the broader category of 'multifunctional materials'. Smart materials are characterized by their significant sensory and actuator capabilities. Sensory capabilities involve the conversion of a mechanical signal into a non-mechanical one, whereas actuation involves the transformation of a non-mechanical input into a mechanical output. In the case of smart materials, this transfer is characterized by a high sensitivity, denoting a substantial change in output for a small alteration in input, which is an important feature for a sensor or actuator \cite{lagoudas2008shape}. The coupling effect underlying these sensing and actuating capabilities in shape memory alloys is based on thermomechanics. In the case of shape memory alloys, both mechanical and thermal fields can serve as input parameters, so that the other field acts as the output parameter. This bidirectional coupling is known as direct coupling and is a distinguishing feature among active materials.
The unique thermomechanical coupling gives rise to specific phenomena referred to as pseudoelasticity, the one-way effect, and the two-way effect. Further information on phase transformation processes and the resulting effects mentioned can be found in \cite{otsuka2005physical}, \cite{shaw1995thermomechanical}, and \cite{lagoudas2008shape}. These distinctive attributes make shape memory alloys exceptionally well-suited for diverse applications, encompassing aerospace \cite{bil2013wing}, medical \cite{petrini2011biomedical}, automotive \cite{jani2014shape} and robotics \cite{sreekumar2007critical} fields. 
Given the complex material properties of shape memory alloys, which are strongly influenced by thermal and mechanical loading conditions as well as loading history, predicting the material response is challenging. Consequently, there is a significant industrial demand for a material model that enables a realistic simulation of components made from shape memory alloys. However, for practical application, it is essential that the corresponding material model has acceptable computational efficiency. For this reason, the description of the present material model is based on a macroscopic level, which offers advantages in terms of computational effort.

\noindent In the literature, several material models for shape memory alloys are described at the macroscopic level. These models utilize various theoretical approaches and fundamental principles to capture the complex behavior of these alloys. 
A one-dimensional phenomenological model that focuses on tension-compression asymmetries and phase transformation-dependent elastic properties is described e. g. in \cite{auricchio2009macroscopic}. In \cite{boyd1996thermodynamical}, a thermodynamic model is presented that can predict both the pseudoelasticity effect and the shape memory effect. In \cite{lagoudas2012constitutive}, a thermomechanical constitutive model based on \cite{boyd1996thermodynamical} is presented that includes important aspects of SMA behavior that were not addressed in the existing models.
In \cite{lagoudas2006shape}, macroscopic approaches for polycrystalline shape memory alloys are discussed, which include the validation of various models using experimental data and the investigation of rate-independent phenomenological models. These models specifically address the inelastic behavior associated with phase transformations and transformation-induced plasticity. Further relevant macro-level material models for shape memory alloys are proposed in \cite{chemisky2011constitutive}, \cite{arghavani20103}, \cite{reese2008finite}, \cite{scalet2019three}, \cite{mohan2021modelling}, \cite{xu2019three} and \cite{xu2021finite}, which focus on different aspects.
A particularly noteworthy macro-level material model was introduced by Junker et al. in \cite{junker2014accurate}. This model stands out due to its solid theoretical framework and was extended to incorporate the concept of plasticity in \cite{junker2017numerical}. Remarkably, it was demonstrated in \cite{JUNKER201586} that this model reliably predicts the mechanical behavior of shape memory alloys in tensile tests, even though most of the model parameters were determined exclusively from thermal experiments rather than mechanical tests. This result indicates high quality of the basic principle on which the material models in \cite{junker2014accurate}, \cite{JUNKER201586} and \cite{junker2017numerical} are based.

\noindent Given this high quality, the current material model is also based on the fundamental principle known as Hamilton's principle. Nevertheless, this material model offers advantages over other models based on the same principle. One of the benefits of the presented material model is the rate-independent evolution of the internal variables, which are crucial for the accurate representation of the phase transformation. This property is particularly relevant as phase transformations in shape memory alloys are inherently diffusionless \cite{otsuka1999shape}, making it time-independent—or in other terms, rate-independent. Despite this property, rate-dependent evolution equations have been used in some other works, for instance, in \cite{junker2017numerical}, \cite{junker2016calibration} and \cite{junker2014accurate}. The use of rate-independent evolution equations is enabled by algorithms described in section 3.
Another significant advantage is the much higher robustness of the present material model, as evidenced by its convergence on complex boundary value problems. The increased robustness results from the correction of the analytically computed Lagrange multiplier, which become inaccurate due to the time discretization. The correction is done by algorithms described in section 3. They constitute an easy-to-implement alternative compared to the exponential mapping time integration method according to \cite{weber1990finite}, which also ensures that constraints are met. In e.g. \cite{reese2008finite}, this method was used for the time discretization of the evolution equations.
A further advantage of the present material model is the parametrization of the rotation matrix using Euler-Rodrigues parameters instead of Euler angles, since Euler-Rodrigues parameters are not susceptible to the Gimbal Lock issue. Gimbal Lock is a phenomenon where two of the three rotational axes in a three-axis system become aligned, resulting in the loss of one degree of rotational freedom.
Another advantage relates to the calculation method of the volume fractions of the crystallographic phases. In the present work, we refrain from using the active set method due to its potentially slow convergence rate for certain types of problems, especially when the active set is changed frequently. Moreover, it is time-consuming to implement the active set method. As an alternative, a penalty term is incorporated, and the internal variable relating to the volume fractions is substituted using the sigmoid function. This also ensures that constraints are met and also improves the convergence performance.
The last advantage refers to the representation of twinned martensite. In \cite{junker2017numerical}, the material response differs depending on the loading direction when twinned martensite is represented through the internal variables. However, twinned martensite inherently exhibits a uniform material response across all loading directions. To model this more accurately in the current material framework, the initial values for the Euler-Rodrigues parameters are selected according to a principle explained in section 3. This principle ensures a uniform material response at each integration point, regardless of the loading direction. An accurate representation of twinned martensite is essential, for example, to correctly simulate the characteristic one-way effect.

\noindent Concluding, it can be stated that the developed material model represents an important continuation of previous research and thus provides deeper insights into the understanding of the behavior of shape memory alloys (SMA). 
This improved understanding enables a more accurate characterization of the unique behavior of SMA. Thus, the model can serve as a basis for innovative applications in various industries.

\noindent This paper is divided into six sections. Following the introduction, the second section introduces basic equations and derives the evolution equations.
The subsequent section offers a comprehensive overview on the structure of the UMAT subroutine. In this context, improvements of the implementation are examined and specific algorithms are presented that improve the robustness of the material model. The fourth section shifts the focus to the computation of quantities that must be supplied to the finite element software Abaqus within the UMAT subroutine.
The penultimate section refers to the numerical results. In this section, investigations are carried out on the material model. First, the influence of the number of load steps on the material response is investigated in a tensile test. Subsequently, a mesh convergence study is performed on a plate with a hole. Then, two complex boundary value problems from the field of medical technology are presented. The simulation of a stent under manufacturing and application conditions represents the first boundary value problem.  The other boundary value problem refers to the simulation under manufacturing and use conditions of a staple used to treat bone fractures. 
The final section summarizes the key findings and draws conclusions based on the research presented.

\section{Material Model}\label{sec2}
The derivation of the material model is based on Hamilton's principle. The Hamilton principle states that the path taken by the physical system between two particular times and configurations is the one for which the action is stationary over an arbitrary time interval (\(\tau= t_{1}-t_{0}\)). According to \cite{junker2021extended}, the action for non-conservative continua can be expressed by the extended Hamilton functional, which is simplified to  
\renewcommand{\CancelColor}{\color{mytobi}}
\begin{align} 
\mathcal{H}^{\mathrm{ext}} := &
   \int\limits_{\tau}^{}
\Big(  
     \; \;  - \; \; \mathcal{G} \; \; 
    -\int\limits_{\Omega}^{} \int\limits_{\tau}^{} \int\limits_{}^{}
    \Delta_{\mathrm{diss}} 
    \; \, \mathrm{d}t \, \mathrm{d}t \, \mathrm{d}V 
    -\int\limits_{\Omega}^{} \int\limits_{\tau}^{}
    \pmb{l} \cdot \pmb{c}
    \; \, \mathrm{d}t \, \mathrm{d}V
\Big) \mathrm{d}t \quad .  \label{eq:extendedHam} 
\end{align}
The extended Hamiltonian functional $\mathcal{H}^{\mathrm{ext}}$ is characterized by the dissipation function \(\Delta_{\mathrm{diss}}\) and the total potential $\mathcal{G}$. Furthermore, according to the Lagrange multiplier method, constraints are integrated into the Hamiltonian via the vectors $\pmb{l}$ and $\pmb{c}$. The total potential $\mathcal{G}$ is defined as
\begin{align}
\mathcal{G} :=  \; \mathcal{E}  \; -   \; \mathcal{W}  \;  = \int\limits_{\Omega}^{}  \; \rho^{(0)}  \, \Psi \; \mathrm{d}V \,    
- \int\limits_{\Omega}^{} \mathbf{b}^* \cdot \mathbf{u} \, \mathrm{d}V \, - \int\limits_{\partial\Omega}^{} \mathbf{t}^* \cdot \mathbf{u} \, \mathrm{dA} \quad .   
\end{align} 
This quantity expresses the difference between the energy $\mathcal{E}$, stored in the system, and the work done by the volume forces $\mathbf{b}^*$ as well as the tractions $\mathbf{t}^*$. The internal energy $\mathcal{E}$  is represented by the elastic strain energy, also known as Helmholtz free energy $\Psi$, which is integrated over the region $\Omega$ and multiplied by the density $\rho$.
\noindent Since the Helmholtz free energy $\Psi$ depends on the microstructure described by the internal variables, several internal variables are first introduced. In the case of shape memory alloys, essential characteristics of their microstructure include the volume fractions of crystallographic phases, grain orientations, dislocations, and hardening. To account for these factors, four internal variables are introduced. The first set of variables is the volume fractions of the crystallographic phases. These fractions are represented by the vector \(\pmb{\lambda}=[\lambda_{0},\lambda_{1},\lambda_{2},\lambda_{3}]\), where the index \(i\) corresponds to different variants of the crystal lattice. The index \(i=0\) denotes the austenite phase, whereas the martensite phases are denoted by \(i=1\), \(i=2\) and \(i=3\).
The second set of internal variables accounts for the average reorientation of the crystallographic  martensite phases and is described by a rotation matrix \(\pmb{Q}(\pmb{\alpha})\). Unlike other works in which Euler angles are used, the parameterization of the rotation matrix in this work is done using Euler-Rodrigues parameters denoted by \(\pmb{\alpha}= [a,b,c,d]\).
Euler-Rodrigues parameters are preferred over Euler angles to avoid Gimbal Lock \cite{stuelpnagel1964parametrization}. It should be noted that the use of Euler-Rodrigues parameters requires satisfying a condition, which can be easily enforced using the Lagrange multiplier method. Another internal variable, denoted as \(\pmb{\varepsilon}_{\mathrm{pl}}\), is introduced to account for dislocations and irreversible deformations, respectively. Additionally, the effects of hardening are considered and represented by the variable \(\kappa\). All of the internal variables are comprised by \( \pmb{\xi} = [\pmb{\lambda},\pmb{\alpha},\pmb{\varepsilon}_{\mathrm{pl}},\kappa] \).

\begin{figure}[H]
    \centering
    	\begin{tikzpicture}[scale=1.0]
    \begin{groupplot}[
                group style={       
                group size=1 by 3,
                xticklabels at=edge bottom,
                vertical sep=0pt
            },
            legend cell align=left,
            legend pos=north east,
            width=0.9\textwidth,
            height = 10.5cm,
            xmin=0, 
            xmax=1
        ]
        \nextgroupplot[ymin=49,ymax=50,
                        ytick={49,49.5,...,50},
                        axis x line=top, 
                        height = 3.5cm
        ]
        \legend{
              {$\Lambda=10^{-5}$},
              {$\Lambda=10^{-4}$},
              {$\Lambda=10^{-3}$}
            }
        \addplot[domain=0.99:0.9999, samples=50, red, line width=2pt, restrict y to domain=-10:200] {1e-5/(((x^2)*((1-x)^2)))};
        \addplot[blue,domain=0.99:0.9999, samples=50, line width=2pt,dashed, restrict y to domain=-10:100] {1e-4/(((x^2)*(1-x)^2))}; 
        \addplot[ color=green!60!black,domain=0.99:0.9999, samples=50, line width=2pt,dotted, restrict y to domain=-10:100] {1e-3/(((x^2)*(1-x)^2))}; 
        \addplot[domain=0.0001:0.01, samples=50, red, line width=2pt, restrict y to domain=-10:200] {1e-5/(((x^2)*((1-x)^2)))};
        \addplot[blue,domain=0.0001:0.01, samples=50, line width=2pt,dashed, restrict y to domain=-10:100] {1e-4/(((x^2)*(1-x)^2))}; 
        \addplot[ color=green!60!black,domain=0.001:0.01, samples=50, line width=2pt,dotted, restrict y to domain=-10:100] {1e-3/(((x^2)*(1-x)^2))};         

        \nextgroupplot[ymin=1.0, ymax=2.0,
                        ytick=\empty,
                        hide x axis,
                        height = 2.5cm,
                        axis y discontinuity=crunch]
        \nextgroupplot[ymin=0, ymax=1,
                        ytick={0,0.5,...,1.0},
                        axis x line=bottom,
                        xlabel={\large $\lambda_{i}$},
                        ylabel=\qquad \qquad \qquad \qquad \qquad {\large $\Psi_{\mathrm{pen}, i}$},
                        height = 3.5cm]
\addplot[domain=0.001:0.999, samples=500, red, line width=2pt, restrict y to domain=-10:50] {1e-5/(((x^2)*((1-x)^2)))};
\addplot[domain=0.001:0.999, samples=500, blue, dashed, line width=2pt, restrict y to domain=-10:50] {1e-4/(((x^2)*(1-x)^2))}; 
\addplot[domain=0.001:0.999, samples=500, green!60!black, dotted, line width=2pt, restrict y to domain=-10:50] {1e-3/(((x^2)*(1-x)^2))}; 
        \end{groupplot}
    \end{tikzpicture} 
    \vspace{-2mm}
    \caption{A graphical representation of the artificial energy-volume fraction $\Psi_{\mathrm{pen}, i}$
    correlation with different values for the parameter $\Lambda$}\centering
    \label{fig:artificialEnergy}
\end{figure}
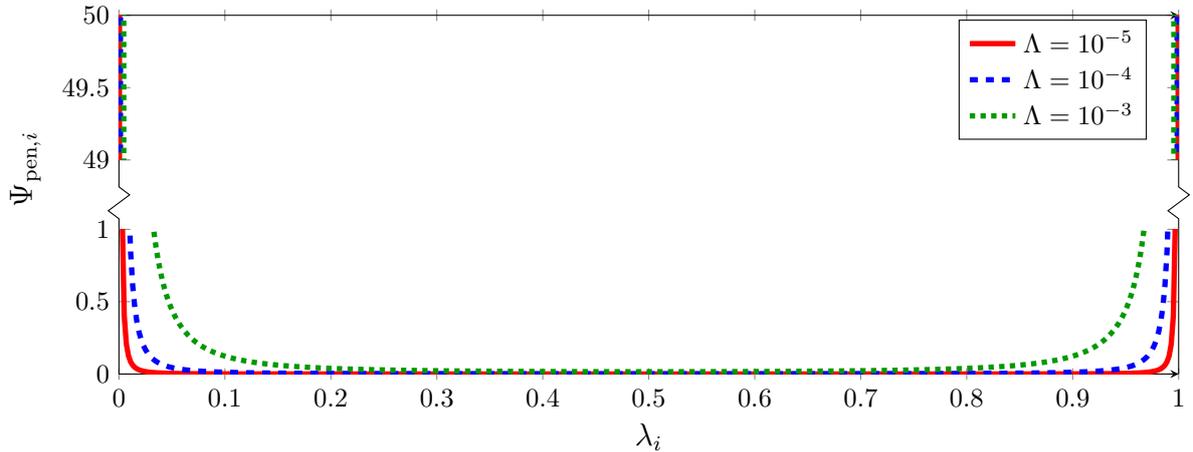

\noindent After introducing the internal variables, the Helmholtz free energy of a crystallographic phase $\Psi_{i}$ can now be expressed mathematically by 
\begin{align}
\Psi_{i}(\pmb{\varepsilon},\theta,\pmb{\lambda},\pmb{\alpha},\pmb{\varepsilon}_{\mathrm{pl}}) 
=\frac{1}{2}\Big(\pmb{\varepsilon} - \pmb{Q}^{T}  \cdot \pmb{\eta_{i}}  \cdot \pmb{Q} - \pmb{\varepsilon}_{\mathrm{pl}}\Big) 
:
{\mathbb{C}_{i}}
:
\Big(\pmb{\varepsilon} - \pmb{Q}^{T}  \cdot \pmb{\eta_{i}}  \cdot \pmb{Q} - \pmb{\varepsilon}_{\mathrm{pl}}\Big) - c_{i}(\theta) \label{eq:freeEnergyPhase}
\end{align}
The first term of the eq. \eqref{eq:freeEnergyPhase} expresses the stored mechanical energy in the corresponding phase. The second term $c_{i}(\theta)$ refers to the energy associated with the thermal motion of the atoms, i.e., the caloric energy.
Here, the total strain is represented by \(\pmb{\varepsilon}\) while the plastic strain is denoted by \(\pmb{\varepsilon}_{\mathrm{pl}}\). The transformation strain of the phase \(i\), labeled as \(\pmb{\eta}_{i}\),
\begin{align}
    \pmb{\eta}_{1} = \hat{\eta}
            \begin{pmatrix}
            1 & 0 & 0  \\
            0 & -\hat{\nu} & 0  \\
            0 & 0 & -\hat{\nu}  \\
            \end{pmatrix}
        \quad  \quad
    \pmb{\eta}_{2} = \hat{\eta}    
            \begin{pmatrix}
            -\hat{\nu} & 0 & 0  \\
            0 & 1 & 0  \\
            0 & 0 & -\hat{\nu}  \\
            \end{pmatrix}
        \quad  \quad
    \pmb{\eta}_{3} = \hat{\eta}
    \begin{pmatrix}
            -\hat{\nu} & 0 & 0  \\
            0 & -\hat{\nu} & 0  \\
            0 & 0 & 1  \\
    \end{pmatrix}
\end{align}
\vspace{-1mm}
are subjected to rotation by the transformation matrix 
\begin{align}
\pmb{Q}(\pmb{\alpha}) = \pmb{Q}(a,b,c,d) =
\begin{pmatrix}
a^{2}+b^{2}-c^{2}-d^{2} & 2(bc -ad) & 2(bd+ac)  \\
2(bc+ad) & a^{2}+c^{2}-b^{2}-d^{2} & 2(cd-ab)  \\
2(bd -ac) & 2(cd + ab) & a^{2}+d^{2}-b^{2}-c^{2} \\ 
\end{pmatrix} \quad.
\end{align}
The parameter \(\hat{\eta}\) represents the scalar maximum transformation strain and \(\hat{\nu}\) is the Poisson's ratio for phase transformation. The symbol \( {\mathbb{C}_{i}} \) refers to the isotropic elasticity tensor of the corresponding phase, which can be calculated using the Young's modulus \(E_{i}\) and the Poisson's ratio \(\nu_{i}\) of the corresponding phase. 
The total free energy in the system is obtained by taking the weighted average of the free energies of the individual phases. For this homogenization, the mixing rule according to Reuss is employed, as it leads to a convex functional unlike the Voigt mixing rule, see \cite{junker2014novel} and \cite{heinen2007calculation}. Applying the mixing rule yields
\begin{equation}
\overline{\Psi}(\pmb{\varepsilon},\theta,\pmb{\lambda},\pmb{\alpha},\pmb{\varepsilon}_{\mathrm{pl}}) = 
\frac{1}{2}\Big(\pmb{\varepsilon} - \pmb{Q}^{T}  \cdot \pmb{\overline{\eta}}  \cdot \pmb{Q} - \pmb{\varepsilon}_{\mathrm{pl}}\Big):
\overline{\mathbb{C}}:
\Big(\pmb{\varepsilon} - \pmb{Q}^{T}  \cdot \pmb{\overline{\eta}}  \cdot \pmb{Q} - \pmb{\varepsilon}_{\mathrm{pl}} \Big) - \overline{c}(\theta)
\end{equation}
with the effective quantities
\begin{align}
\pmb{\overline{\eta}} = \sum_{i=0}^{n} \; \lambda_{i} \pmb{\eta}_{i} \quad;\quad
\overline{c} = \sum_{i=0}^{n} \; \lambda_{i} c_{i}(\theta) \quad;\quad
\overline{\mathbb{C}}= \bigg[ \sum_{i=0}^{n} \;  \lambda_{i} (\mathbb{C}_{i})^{-1} \bigg]^{-1} \quad.
\end{align}
A nonlinear behavior can be observed in tensile tests of nitinol systems, which are used in a broader range and much more frequently in commercial applications compared to other SMA compositions \cite{lagoudas2008shape}, \cite{shaw1995thermomechanical}, \cite{jani2014review}. 
This behavior is attributed to nonlinear isotropic hardening, and it is accounted for by incorporating a specific hardening energy term, denoted as $\Psi_{\mathrm{h}}$, into the total Helmholtz free energy $\Psi$. This total Helmholtz free energy can be expressed as
\begin{equation}
\Psi =  \overline{\Psi} + {\Psi}_{\mathrm{h}}  + {\Psi}_{\mathrm{pen}} \quad .  \label{eq:totalHfE}\\
\end{equation}
The formulation of the hardening energy is given by
\begin{equation}
\Psi_{\mathrm{h}} = \frac{1}{2} \, k_{1} \, \kappa^{2} -
\frac{k_{1}-k_{0}}{k_{2}}\bigg(\frac{1}{k_{2}} e^{-k_{2}\kappa} + \kappa \bigg) \quad .
\end{equation}
Here, \(k_{1}\), \(k_{2}\) and \(k_{3}\) are parameters that determine the shape of the hardening curve. 
Furthermore, eq. \eqref{eq:totalHfE} contains an additional term ${\Psi}^{\mathrm{pen}}$, which is defined as
\begin{equation}
{\Psi}_{\mathrm{pen}} = 
\sum_{i=0}^{n}\frac{\Lambda}{\lambda^{2}_{i}(1 - \lambda_{i})^{2}} \quad . \label{eq:penaltyEnergy}
\end{equation}
\noindent This additional term, called penalty term, acts as an artificial energy that penalizes volume fractions as they approach the extremes of 1 and 0, as illustrated in Fig. \ref{fig:artificialEnergy}. The penalty term in combination with the substitution of the internal variable $\lambda$ using a sigmoid function ensures that the inequality condition $0 \leq \lambda \leq 1$ is fulfilled. In the further course of the section, this aspect will be revisited. The parameter \(\Lambda\) influences the magnitude of the added artificial energy.
\vspace{-7mm}
\vspace{-6pt}
\noindent The term $-\int\limits_{\Omega}^{} \int\limits_{\tau}^{} \int\limits_{}^{}
\Delta_{\mathrm{diss}} \, \mathrm{d}t \, \mathrm{d}t \, \mathrm{d}V $ of the extended Hamilton functional eq. \eqref{eq:extendedHam}, accounts for the energy dissipated within the microstructure. The choice of dissipation function $\Delta_{\mathrm{diss}}$ determines whether the internal variables undergo rate-dependent or rate-independent evolution. More details can be found in \cite{junker2021extended}.\\
In this work, a rate-independent approach is chosen for all internal variables, attributed to the diffusionless nature of the transformation process in shape memory alloys  \cite{lagoudas2008shape}. Therefore, it is expected that the internal variables evolve instantaneously rather than exhibiting a time delay. Thus, the dissipation function is defined as  
\begin{equation}
\Delta_{\mathrm{diss}} = r_{\lambda}||\dot{\pmb{\lambda}}||+r_{\alpha}||\dot{\pmb{\alpha}}||+r_{\mathrm{pl}}||\dot{\pmb{\varepsilon}}_{\mathrm{pl}}|| \quad .
\end{equation}
\noindent 
As mentioned before the Lagrange multiplier method is used to restrict the solution set as desired. The scalar product of the vectors $\pmb{c}$ and $\pmb{l}$ is formed for this purpose.
The vector $ \pmb{c}$ includes constraints rearranged to zero, and the vector $ \pmb{l} $ comprises the multipliers, specifically $ \gamma $, $ \zeta $, $ \beta $, and $ \mu $.
Two of the constraints relating to volume conservation apply to the volume fraction of the crystallographic phases and the plastic strains, which must fulfill the constraints
\vspace{-3mm}
\begin{align}
\sum_{i=0}^{n}\lambda_{i} = 1 \qquad &\Longleftrightarrow \qquad c_{\lambda}:= \sum_{i=0}^{n}\lambda_{i} -1 = 0 \label{eq:consLambda} 
\\
 &\text{and} \nonumber
\\
c_{\varepsilon_{\mathrm{pl}}}:= \sum_{j=1}^{3}&\varepsilon_{\mathrm{pl},jj} = 0 \quad . \label{eq:consEpspl}
\end{align}
In this work, Euler-Rodrigues parameter are used, for which the condition \( \pmb{\alpha} \cdot \pmb{\alpha} = 1 \) must be satisfied \cite{stuelpnagel1964parametrization}. This leads to the following constraint
\begin{equation}
\pmb{\alpha} \cdot \pmb{\alpha} = 1 \qquad \Longleftrightarrow \qquad 
c_{\alpha}:= (\pmb{\alpha} \cdot \pmb{\alpha} - 1)  = 0 \quad . \label{eq:consAlpha}
\end{equation}
Additionally, a classical assumption is made that \( \dot{\kappa} = ||\dot{\pmb{\varepsilon}}_{\mathrm{pl}}|| \), resulting in the constraint condition
\begin{equation}
\dot{c}_{\kappa}:=
||\dot{\pmb{\varepsilon}}_{\mathrm{pl}}|| - \dot{\kappa} = 0  \qquad \Longleftrightarrow \qquad  c_{\kappa}:=
\int\limits_{\tau} ||\dot{\pmb{\varepsilon}}_{\mathrm{pl}}|| \, \mathrm{d}t - \kappa = 0 . \\ \label{eq:consAlphaH}
\end{equation}
Consequently, the dot product of \( \pmb{l} \cdot \pmb{c}\) is expressed as 
\begin{align}
\eta \; := \; \pmb{l} \cdot \pmb{c} \; = \gamma \Big( \pmb{1}_{\mathrm{n}} \cdot \pmb{\lambda} - 1\Big) + \beta \Big( \pmb{\alpha} \cdot \pmb{\alpha} - 1 \Big) + \zeta \Big( \pmb{I}_{3}:\pmb{\varepsilon}_{\mathrm{pl}} \Big) + \mu \Big(||\dot{\pmb{\varepsilon}}_{\mathrm{pl}}|| - \dot{\kappa} \Big) \; . \label{eq:cons}
\end{align}
Here, $\pmb{1}_\mathrm{n}$ is a vector of length $n$ with all entries being ones. As mentioned earlier, Hamilton's principle states that the action integral introduced in eq. \eqref{eq:extendedHam} has to become stationary, which means that
\begin{equation}
 \mathcal{H} \rightarrow \underset{\pmb{\xi}}{\mathrm{stat}} \qquad \Longleftrightarrow \qquad \delta \mathcal{H}[\pmb{u},\pmb{\xi}] (\delta \,\pmb{u} , \delta \,\pmb{\xi}) = 0 \quad \forall \; \delta \,\pmb{u} , \delta \pmb{\xi} \quad . 
\end{equation}
The evaluation of the stationary condition with respect to the internal variables, using the Gateaux differential, leads to
\begin{equation}
    - \;
    \frac
        {\partial \, \Psi }{\partial \, \pmb{\xi}} \; \; = \; \; \hat{\pmb{p}}_{\Delta} + \, \hat{\pmb{p}}_{\mathrm{c}}
    \;    \quad .
    \label{eq:reduced_lagrange_equation}
\end{equation}
The non-conservative forces are defined as  $\hat{\pmb{p}}_{\mathrm{c}} := \frac{\partial \, \eta}{\partial \, \pmb{\xi}}$ and $\hat{\pmb{p}}_{\Delta} := \frac{\partial  \int\limits \Delta_{\mathrm{diss}} \, \mathrm{d}t}{\partial \, \pmb{\xi}}$. Applying the chain rule leads to $\frac{\partial  \int\limits \Delta_{\mathrm{diss}} \, \mathrm{d}t}{\partial \, \pmb{\xi}} = \frac{\partial \, \Delta_{\mathrm{diss}}}{\partial \, \dot{\pmb{\xi}}}$.
\noindent  
The non-conservative force based on the dissipation within the microstructure \( \hat{\pmb{p}}_{\Delta}\) can be expressed in component notation as
\begin{align}
    \hat{\pmb{p}}_{\Delta,\lambda} := \; \; \frac{\partial \, \Delta_{\mathrm{diss}}}{\partial \dot{\pmb{\lambda}}} \,  = \; & r_{\lambda}\frac{\dot{\pmb{\lambda}}}{||\dot{\pmb{\lambda}}||}
    \label{eq:ddiss_dlambda} \\
    \hat{\pmb{p}}_{\Delta,\alpha} := \; \; \frac{\partial \, \Delta_{\mathrm{diss}}}{\partial \dot{\pmb{\alpha}}} \,  = \; &  r_{\alpha}\frac{\dot{\pmb{\alpha}}}{||\dot{\pmb{\alpha}}||}
    \label{eq:ddiss_dalpha} \\
    \hat{\pmb{p}}_{\Delta,\varepsilon_{pl}} := \; \; \frac{\partial \, \Delta_{\mathrm{diss}}}{\partial \dot{\pmb{\varepsilon}}_{\mathrm{pl}}} \, =  \; & r_{\mathrm{pl}}\frac{\dot{\pmb{\varepsilon}}_{\mathrm{pl}}}{||\dot{\pmb{\varepsilon}}_{\mathrm{pl}}||}
    \label{eq:ddiss_deps_pl} \\
    \hat{p}_{\Delta,\kappa} := \; \; \frac{\partial \, \Delta_{\mathrm{diss}}}{\partial \dot{\kappa}} \, 
    = \; & 0 \qquad . \label{eq:ddiss_dkappa} 
\end{align}
\vspace{-5mm}
And the non-conservative force based on the constraints \( \hat{\pmb{p}}_{\mathrm{c}}\) can be represented in component form as follows:
\vspace{3mm}
\begin{alignat}{2}
    \hat{\pmb{p}}_{\mathrm{c},\lambda} &:= \; \; \frac{\partial \, \eta}{\partial \, \pmb{\lambda}} &&=
    \frac{\partial \, c_{\lambda}}{\partial \, \pmb{\lambda}}
    =\gamma \,\pmb{1}_{n}
    \label{eq:dcons_dlambda} \\
    \hat{\pmb{p}}_{\mathrm{c},\alpha} &:= \; \; \frac{\partial \, \eta}{\partial \, \pmb{\alpha}} &&=
    \frac{\partial \, c_{\lambda}}{\partial \, \pmb{\alpha}}=\beta \, 2\pmb{\alpha}
    \label{eq:dcons_dalpha} \\
\hat{\pmb{p}}_{\mathrm{c},\varepsilon_{\mathrm{pl}}} &:= \; \; \frac{\partial \, \eta}{\partial \, \pmb{\varepsilon}_{\mathrm{pl}}} &&=%
\frac{\partial \, c_{\varepsilon_{\mathrm{pl}}}}{\partial \, \pmb{\varepsilon}_{\mathrm{pl}}} + \frac{\partial \, c_{\kappa}}{\partial \, \pmb{\varepsilon}_{\mathrm{pl}}}
=
\frac{\partial \, c_{\varepsilon_{\mathrm{pl}}}}{\partial \, \pmb{\varepsilon}_{\mathrm{pl}}} + \frac{\partial \, \dot{c}_{\kappa}}{\partial \, \dot{\pmb{\varepsilon}}_{\mathrm{pl}}}
= \zeta \, \pmb{I}_{3} + \mu \frac{\dot{\pmb{\varepsilon}}_{\mathrm{pl}}}{||\dot{\pmb{\varepsilon}}_{\mathrm{pl}}||}
    \label{eq:dcons_depsilon_pl} \\
    \hat{\pmb{p}}_{\mathrm{c},\kappa} &:= \; \; \frac{\partial \, \eta}{\partial \, \kappa} &&= \frac{\partial \, c_{\kappa}}{\partial \, \kappa} = - \mu  \quad .
    \label{eq:dcons_dkappa}
\end{alignat}
\noindent The partial derivative of the total Helmholtz free energy ${\Psi}$ with respect to the internal variables
\( \pmb{\xi} = \pmb{\xi}(\pmb{\lambda},\pmb{\alpha},\pmb{\varepsilon}_{\mathrm{pl}},\kappa) \) yields
\begin{align}
    \frac{\partial \, \Psi}{\partial \, \lambda_{i}}  =&     
        \; \frac{\Lambda(2 - 4\lambda_{i})}{\lambda^{3}_{i}(-1 + \lambda_{i})^{3}} +c_{i}(\theta) -(Q^{T} \cdot \eta_{i} \cdot Q) : \overline{\mathbb{C}} : 
        (\pmb{\varepsilon} - \pmb{Q}^{T}  \cdot \pmb{\overline{\eta}}  \cdot \pmb{Q} - \pmb{\varepsilon}_{\mathrm{pl}})  \qquad \\
        &- \frac{1}{2}(\pmb{\varepsilon} - \pmb{Q}^{T}  \cdot \pmb{\overline{\eta}}  \cdot \pmb{Q} - \pmb{\varepsilon}_{\mathrm{pl}}) \nonumber
        :\big[\overline{\mathbb{C}}:(\mathbb{C}_{i})^{-1}:\overline{\mathbb{C}}\big]:(\pmb{\varepsilon} - \pmb{Q}^{T}  \cdot \pmb{\overline{\eta}}  \cdot \pmb{Q} - \pmb{\varepsilon}_{\mathrm{pl}}) \nonumber \\ 
        \frac{\partial \, \Psi}{\partial \, \alpha_{i}}  =&
        \label{eq:dpsi_dalpha1}
        -2  \Bigl[ \overline{\pmb{\eta}} \cdot Q \cdot          
            \overline{\mathbb{C}} : 
            (
            \pmb{\varepsilon} - \pmb{Q}^{T} \cdot \pmb{\overline{\eta}}            \cdot \pmb{Q} -\pmb{\varepsilon}_{\mathrm{pl}}
            ) 
    \Bigr]: \frac{\partial \pmb{Q}}{\partial \alpha_{i}} \\
    \frac{\partial \, \Psi}{\partial \, \pmb{\varepsilon}_{\mathrm{pl}}}  =&
        \label{eq:dpsi_depspl}
        - \overline{\mathbb{C}} : 
            (
            \pmb{\varepsilon} - \pmb{Q}^{T} \cdot \pmb{\overline{\eta}}            \cdot \pmb{Q} -\pmb{\varepsilon}_{\mathrm{pl}}
            ) \\
    \frac{\partial \, \Psi}{\partial \, \kappa}  =&
        \label{eq:dpsi_dkappa}
 \; k_{1} \, \kappa +
\frac{k_{1}-k_{0}}{k_{2}}\bigg(e^{-k_{2}\kappa} - 1 \bigg) \quad .
\end{align}
\noindent By rearranging eq. \eqref{eq:reduced_lagrange_equation} and inserting the constraints from eq. \eqref{eq:consLambda}-\eqref{eq:consAlpha}, the evolution equations are obtained as
\begin{align}
&\dot{\pmb{\lambda}} \neq \;  \pmb{0}: \quad
&&\dot{\pmb{\lambda}} = \frac{||\dot{\pmb{\lambda}}||}{r_{\lambda}}
    \left[
        -\frac{\partial\Psi}{\partial\pmb{\lambda}} + \pmb{1}_{\mathrm{n}}\frac{1}{n}\sum_{i=0}^{n}\frac{\partial\Psi}{\partial\lambda_{i}}\lambda_{i} 
    \right] &&\hspace{-0.5cm}:= \rho_{\lambda} \; \overline{\pmb{p}}_{\lambda} \label{eq:eveq_lambda}
\\
&\dot{\pmb{\alpha}} \neq \;  \pmb{0}: \quad
&&\dot{\pmb{\alpha}} = \frac{||\dot{\pmb{\alpha}}||}{r_{\alpha}} 
    \left[-\frac{\partial\Psi}{\partial\pmb{\alpha}} + \frac{\partial\Psi}{\partial\pmb{\alpha}} \cdot       \pmb{\alpha} \; \pmb{\alpha}\right] &&\hspace{-0.5cm}:= \rho_{\alpha} \; \overline{\pmb{p}}_{\alpha}
\\
&\dot{\pmb{\varepsilon}}_{\mathrm{pl}} \neq \;  \pmb{0}: \quad 
&&\dot{\pmb{\varepsilon}}_{\mathrm{pl}} =  \frac{||\dot{\pmb{\varepsilon}}^{\mathrm{pl}}||}{r_{\mathrm{pl}} + \mu }
    \left[-\frac{\partial\Psi}{\partial\pmb{\varepsilon}_{\mathrm{\mathrm{pl}}}} + \frac{1}{3} \frac{\partial\Psi}{\partial\pmb{\varepsilon}_{\mathrm{pl}}}:\pmb{I} \; \pmb{I}\right]&&\hspace{-0.5cm}
:= \rho_{\mathrm{pl}} \; \mathrm{dev} \,\pmb{\sigma} \quad  \label{eq:eveq_kappa}
\end{align}
with the consistency parameters $\rho_{\lambda}:=\frac{||\dot{\pmb{\lambda}}||}{r_{\lambda}}$, $\rho_{\alpha}:=\frac{||\dot{\pmb{\alpha}}||}{r_{\alpha}}$ and $\rho_{\mathrm{pl}}:=\frac{||\dot{\pmb{\varepsilon}}^{\mathrm{pl}}||}{r_{\mathrm{pl}} + \mu }$.
The discretization of the evolution eq. \eqref{eq:eveq_lambda} using explicit or implicit time integration methods gives rise to a risk of violating the constraint \( 0 \leq \lambda_{i} \leq 1 \).
leads to unphysical behaviour. To address this issue, the volume fractions \(\pmb{\lambda}\) are expressed in terms of the internal variable \( \pmb{\chi}\) using the sigmoid function 
\begin{align}
 \lambda_{i}(\chi_{i}) = \frac{1}{1 + e^{-\chi_{i}}} \quad . 
\end{align}
The sigmoid function ensures that the volume fractions remain within the range \( 0 \leq \lambda_{i} \leq 1 \). Consequently, the evolution eq. \eqref{eq:eveq_lambda} can be extended to 
\begin{align}
\frac{\mathrm{d} \pmb{\lambda}}{\mathrm{d} t} = 
\frac{\partial\pmb{\lambda}}{\partial \pmb{\chi}} \;
\frac{\partial \pmb{\chi}}{\partial t}
\qquad \Longleftrightarrow \qquad
\dot{\pmb{\chi}} =\bigg(\frac{\partial\pmb{\lambda}}{\partial\pmb{\chi}}\bigg)^{-1} \dot{\pmb{\lambda}} \quad .
\end{align}
\noindent The evolution equations from eq. \eqref{eq:eveq_lambda}-\eqref{eq:eveq_kappa} are active only in case of evolution, i.e. \(||\dot{\pmb{\lambda}}||\neq 0\), \(||\dot{\pmb{\alpha}}||\neq 0\) and \(||\dot{\pmb{\varepsilon}}_{\mathrm{pl}}|| \neq 0\), as set-valued expressions are present at stationarity points, i.e. \(||\dot{\pmb{\lambda}}||= 0\), \(||\dot{\pmb{\alpha}}||= 0\) and \(||\dot{\pmb{\varepsilon}}_{\mathrm{pl}}|| = 0\). 
\noindent Specific criteria, corresponding to the well-known flow function can be obtained for each internal variable by analyzing the Legendre transformation of the dissipation function. A comprehensive derivation is shown in the appendix B, which provides the following criteria
\begin{align}
    &&\dot{\pmb{\lambda}} = \;  \pmb{0}: \qquad
    \phi_{\lambda} :=& \;
    ||\overline{\pmb{p}}_{\lambda}|| - r_{\lambda} \leq 0
    \label{eq:lambda_ff} \\
    &&\dot{\pmb{\alpha}} = \;  \pmb{0}: \qquad
    \phi_{\alpha} :=& \;
    ||\overline{\pmb{p}}_{\alpha}|| - r_{\alpha} \leq 0
    \label{eq:alpha_ff} \\
    &&\dot{\pmb{\varepsilon}}_{\mathrm{pl}} = \;  \pmb{0}: \qquad
    \phi_{\mathrm{pl}} :=& \;
    ||\mathrm{dev}\,\pmb{\sigma}|| - (r_{\mathrm{pl}} + \mu) \leq 0 \quad .
    \label{eq:epspl_ff} 
\end{align} 
\noindent The evaluation of the stationary condition of the Hamilton functional with respect to the hardening variable $\kappa$ provides the expression $\mu= \frac{\partial \, \Psi}{\partial \, \kappa}$.
\newpage
\section{Algorithmic Implementation}\label{Algorithmic_Implementation}
\begin{figure}[H]%
\centering
\includegraphics[width=0.75\textwidth]{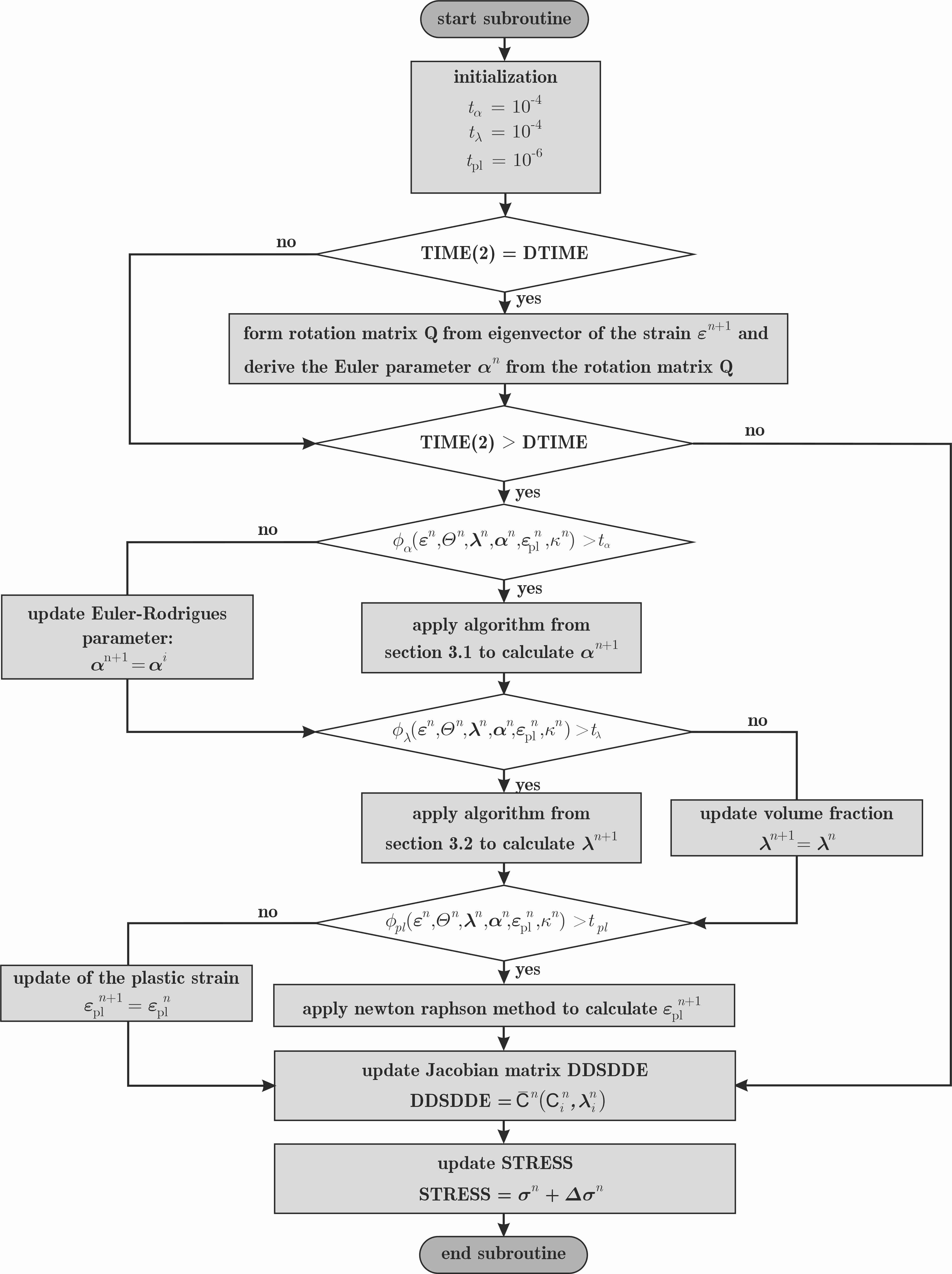}
\caption{Principle structure of the user subroutine UMAT}\label{fig:flowCharttotal}
\end{figure}

\begin{figure}[H]%
\centering
\includegraphics[width=0.35\textwidth]{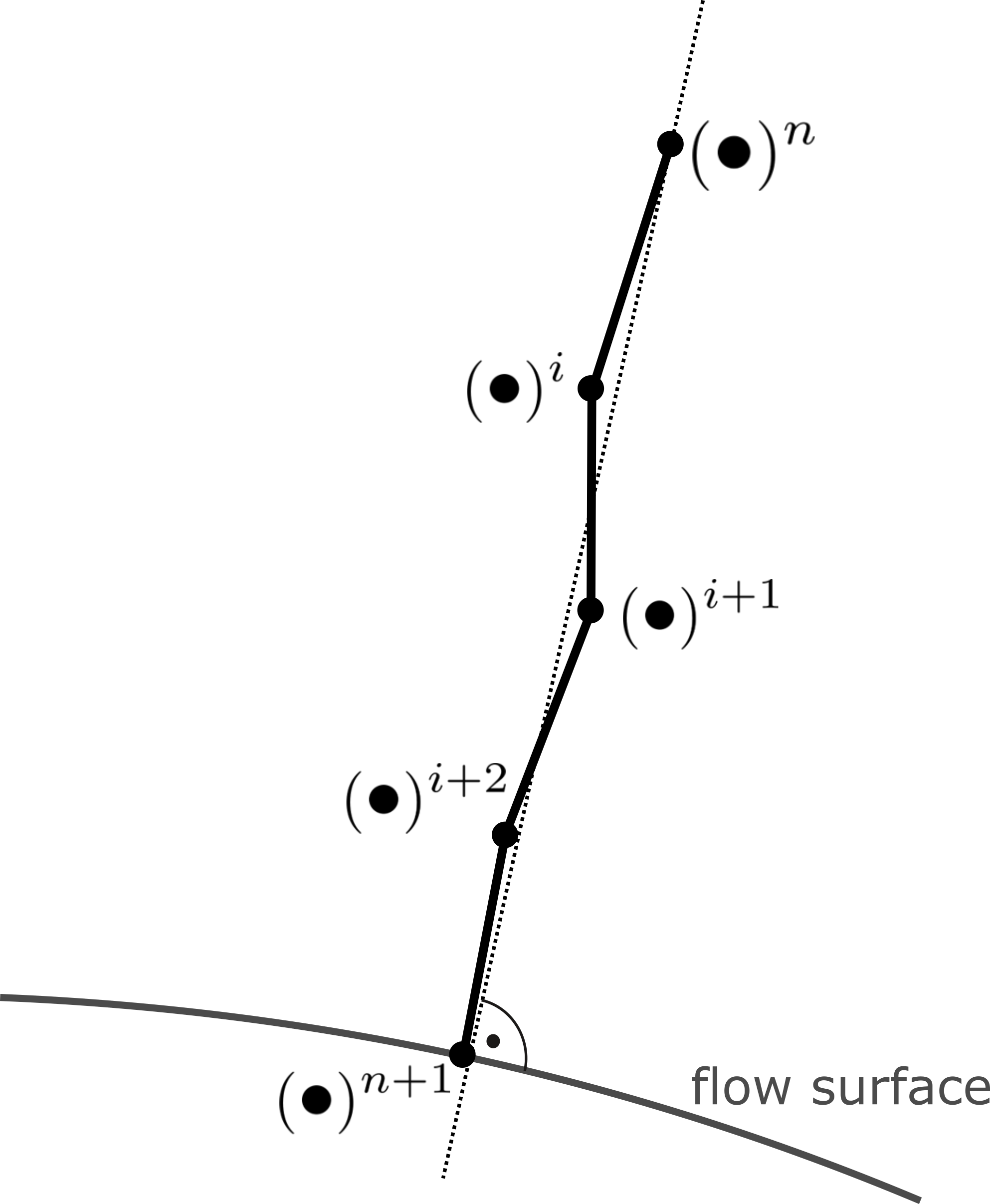}
\caption{Functionality of the algorithms for determining the internal variable in the current time step $(\bigcdot)^{n+1}$}\label{fig:FuntionalityOfTheAlgorithms}
\end{figure}

To demonstrate that the model can be implemented in industrial software environments, we present an example of its algorithmic implementation into the software Abaqus.
Figure \ref{fig:flowCharttotal} illustrates the overall structure of the user subroutine UMAT in Abaqus, represented as a flowchart. Following initialization, the subroutine checks for the presence of the first load increment. This is accomplished by comparing the total time \( \mathrm{TIME(2)}\) with the time increment \( \mathrm{DTIME} \). When these values are equal, it signifies the occurrence of the first load increment. Consequently, the well-known Jacobi method is applied to the strain tensor $\pmb{\varepsilon}^{\mathrm{n+1}}$ to calculate its eigenvectors. This method yields the initial rotation matrix $\pmb{Q}$ 
whose column vectors represent the normalized eigenvectors of the strain tensor.
As a result, the transformation strains \(\pmb{\eta}_{i}\) of the three martensite variants point in the direction of the principal strains, since the transformation strains are rotated using the rotation matrix. The initial Euler-Rodrigues parameters, $\pmb{\alpha}^{0}$, are then derived from this rotation matrix according to \cite{spurrier1978comment}.
If these Euler-Rodrigues parameters are not determined in the stated manner, the microstructure exhibits favored directions for the martensite phase evolution, leading to a material response dependent on the loading direction. This property is usually not present in polycrystalline shape memory alloys. Thus, no arbitrary choice is made for the initial values in this work. 
\vspace{-6pt}
\vspace{-18pt}
\noindent 
The subroutine further checks whether the yield functions $\phi_{\alpha}$, $\phi_{\lambda}$ and $\phi_{\mathrm{pl}}$ exceed their respective tolerance parameters $t_{\alpha}$, $t_{\lambda}$ and $t_{\mathrm{pl}}$.
When these conditions are satisfied, algorithms are applied to compute $\pmb{\alpha}^{n+1}$, $\pmb{\lambda}^{n+1}$, and $\pmb{\varepsilon}_{\mathrm{pl}}^{n+1}$. The classical Newton-Raphson method is used to calculate the plastic strain at the new time step $\pmb{\varepsilon}_{\mathrm{pl}}^{n+1}$. However, direct application of this method to calculate $\pmb{\alpha}^{n+1}$ and $\pmb{\lambda}^{n+1}$ is not feasible because the solution does not converge. 
The non-convergence arises from the analytically computed Lagrange multipliers, which are imprecise as a result of time integration. To solve this problem, special algorithms are presented in sections 3.1 and 3.2. Figure \ref{fig:FuntionalityOfTheAlgorithms} shows schematically how the algorithms work. At the beginning, the thermodynamic state, which depends on the internal variables of the previous time step $(\bigcdot)^{n}$, is outside the flow surface. During each iteration $i$, an update of the internal variable $(\bigcdot)^{n}$ is performed using a Newton-Raphson step. As a result, the thermodynamic state approaches the yield surface. The orientation in each iteration is nearly orthogonal to the yield surface due to a small inaccuracy in the Lagrange multiplier. This discrepancy is subsequently corrected in each iteration. Once the thermodynamic state matches the yield surface, it is accepted as the solution for the current increment $(\bigcdot)^{n+1}$.
A more detailed explanation is given in subsections 3.1 and 3.2.
\vspace{-5mm}
\noindent For the plastic strain, the analytically calculated Lagrange multiplier is also slightly inaccurate. However, the consequences are not as severe as for the Euler-Rodrigues parameter and the volume fractions, since the plastic strains evolve less than the Euler-Rodrigues parameter and the volume fractions. Therefore, the solution for plastic strain still achieves convergence, albeit with tolerable deviation from the constraint.
\vspace{5mm}
\subsection{Calculation of Euler-Rodrigues paramter parameter}\label{algorithm_rodriguesParameter}
\begin{figure}[H]
\centering
\includegraphics[width=0.9\textwidth]{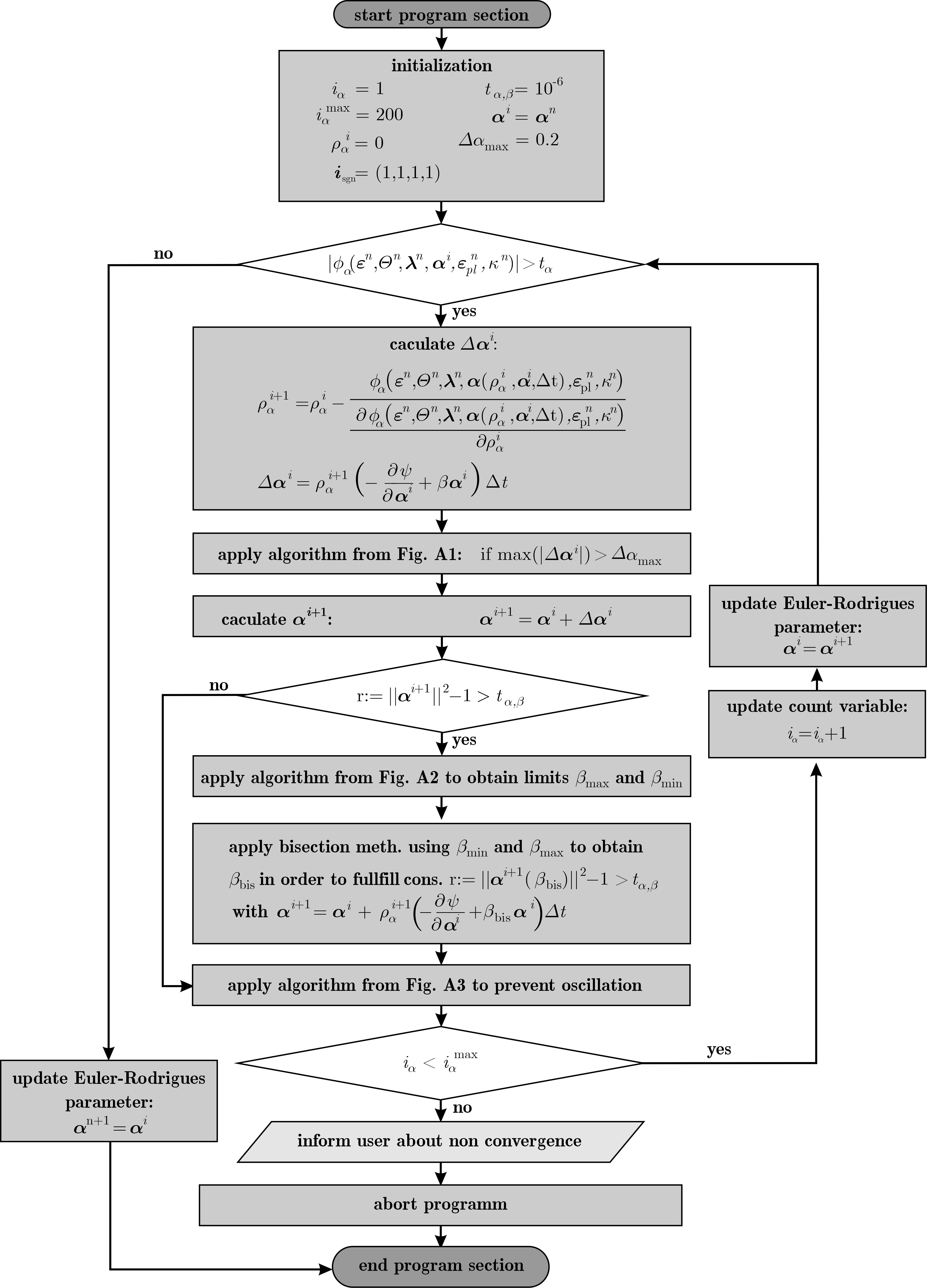}
\vspace{6pt}
\caption{
Algorithm to calculate the Euler parameter $\pmb{\alpha}^{n+1}$} \label{fig:FlowChartAlpha}
\end{figure}

The program section of the UMAT user subroutine in Fig. \ref{fig:FlowChartAlpha} is executed to calculate the Euler-Rodrigues parameter at the new time step $\pmb{\alpha}^{n+1}$. For this purpose, first an initialization is performed. Then, it is checked whether the criterion $|\phi_{\alpha}| > t_{\alpha}$ is fulfilled. If the loop condition $|\phi_{\alpha}| > t_{\alpha}$ is satisfied, the consistency parameter $\rho^{i+1}_{\alpha}$ is computed using one Newton-Raphson step. Consequently, the increment $\Delta\alpha^{i}$ is calculated using the consistency parameter $\rho^{i+1}_{\alpha}$. Note that the Lagrange multiplier and hence the increment $\Delta\alpha^{i}$ is slightly inaccurate as a result of the time discretization, so that the constraint $\pmb{\alpha}^{i+1} \cdot \pmb{\alpha}^{ i+1} = 1 $ is no longer exactly satisfied.
The larger the change in the Euler-Rodrigues parameter, the more the constraint is violated and the more the sought Lagrange multiplier deviates from the analytically calculated one.
A significant violation of the constraint leads to difficulties to set the required limits $\beta_{\mathrm{min}}$ and $\beta_{\mathrm{max}}$, which is necessary to determine the exact Lagrange multiplier using the bisection method.
As a measure, $\Delta\pmb{\alpha}^{i}$ is limited using a scaling factor if any element of the vector $\Delta\alpha^{i}$ is larger than the specified limit $\Delta \alpha_{\mathrm{max,tol}}$. The scaling factor is chosen as shown in Fig. \ref{fig:ProgramSubsectionLimitDeltaAlpha} (appendix) such that the largest element in the vector $\Delta \pmb{\alpha}^{i}$ after scaling, has an absolute value of $\Delta \alpha_{\mathrm{max,tol}}$.
The Euler-Rodrigues parameter are then updated by $ \pmb{\alpha}^{i+1} = \pmb{\alpha}^{i} + \Delta\pmb{\alpha}^{i}$ and the constraint $ || \pmb{\alpha}^{i+1} ||^{2} - 1 = 0 $ is compared to the allowable deviation $ t_{\alpha,\beta}$. If the deviation exceeds the tolerance $ t_{\alpha,\beta} $, bounds $\beta_{\mathrm{min}}$ and $\beta_{\mathrm{max}}$ are defined using the specific algorithm in Fig. \ref{fig:ProgramSubsection_LimitValuesLagrange} (appendix) and the bisection method is used to find the exact Lagrange multiplier $\beta_{\mathrm{bis}}$. The exact Lagrange multiplier $\beta_{\mathrm{bis}}$ is then used to recalculate the Euler-Rodrigues parameter $\pmb{\alpha}^{i+1}$. 
It is worth mentioning that the bisection method has a better performance than the classical Newton-Raphson method in calculating the exact Lagrange multiplier $\beta_{\mathrm{bis}}$.
Moreover, an additional algorithm according to Fig. \ref{fig:ProgramSubsection_preventOscillation} (appendix) is performed to prevent oscillations. 
Further, the iteration $i_{\alpha}$ is evaluated.
If it exceeds $i_{\alpha}^{\mathrm{max}}$, a non-convergence warning is issued and the program terminates. If it is within the limits, iteration continues until the condition $|\phi_{\alpha}| > t_{\alpha}$ is no longer satisfied. At this point, the Euler-Rodrigues parameter from the current iteration $\pmb{\lambda}^{i}$ are accepted as the solution for the current time step $\pmb{\lambda}^{\mathrm{n+1}}$.
\newpage
\subsection{Calculation of volume fractions}\label{algorithm_volumeFraction}
After initialization, a \texttt{while} loop is run as long as the criterion \(|\phi_{\lambda}| > t_{\lambda}\) is satisfied.
If the criterion is unmet, the solution at the current iteration \(i_{\lambda}\) is deemed acceptable, and the volume fractions are consequently updated using \(\pmb{\lambda}^{n+1} \longleftarrow \pmb{\lambda}^{i}\). If the criterion \(|\phi_{\lambda}| > t_{\lambda}\) is satisfied, the consistency parameter $\rho^{n+1}_{\lambda}$ is first determined by a Newton-Raphson step.
Afterwards this consistency parameter $\rho^{n+1}_{\lambda}$ is employed to compute the new volume fractions \(\pmb{\lambda}^{i+1}\).
As mentioned earlier, due to the time discretization, the condition $\sum\lambda^{n+1}_{i} = 1$ is not exactly satisfied, since the analytically determined Lagrange multiplier is not exact. Consequently, the volume fractions are normalized to the value 1 subsequently. Should the permissible number of iterations $i^{\lambda}_{\mathrm{max}}$ be exceeded, the user is notified of the non-convergence of the solution, and the program terminates. If the permissible number of iterations is not surpassed, the iteration counter $i_{\lambda}$ is incremented by one, and the volume fractions are updated as $\pmb{\lambda}^{i} \longleftarrow \pmb{\lambda}^{i+1}$. Subsequently, the termination criterion is evaluated with the updated volume fractions $\pmb{\lambda}^{i}$. If met, the solution for the current time step $n+1$ is deemed acceptable. Otherwise, the loop is repeated.
\begin{figure}[H]%
\centering
\includegraphics[width=0.9\textwidth]{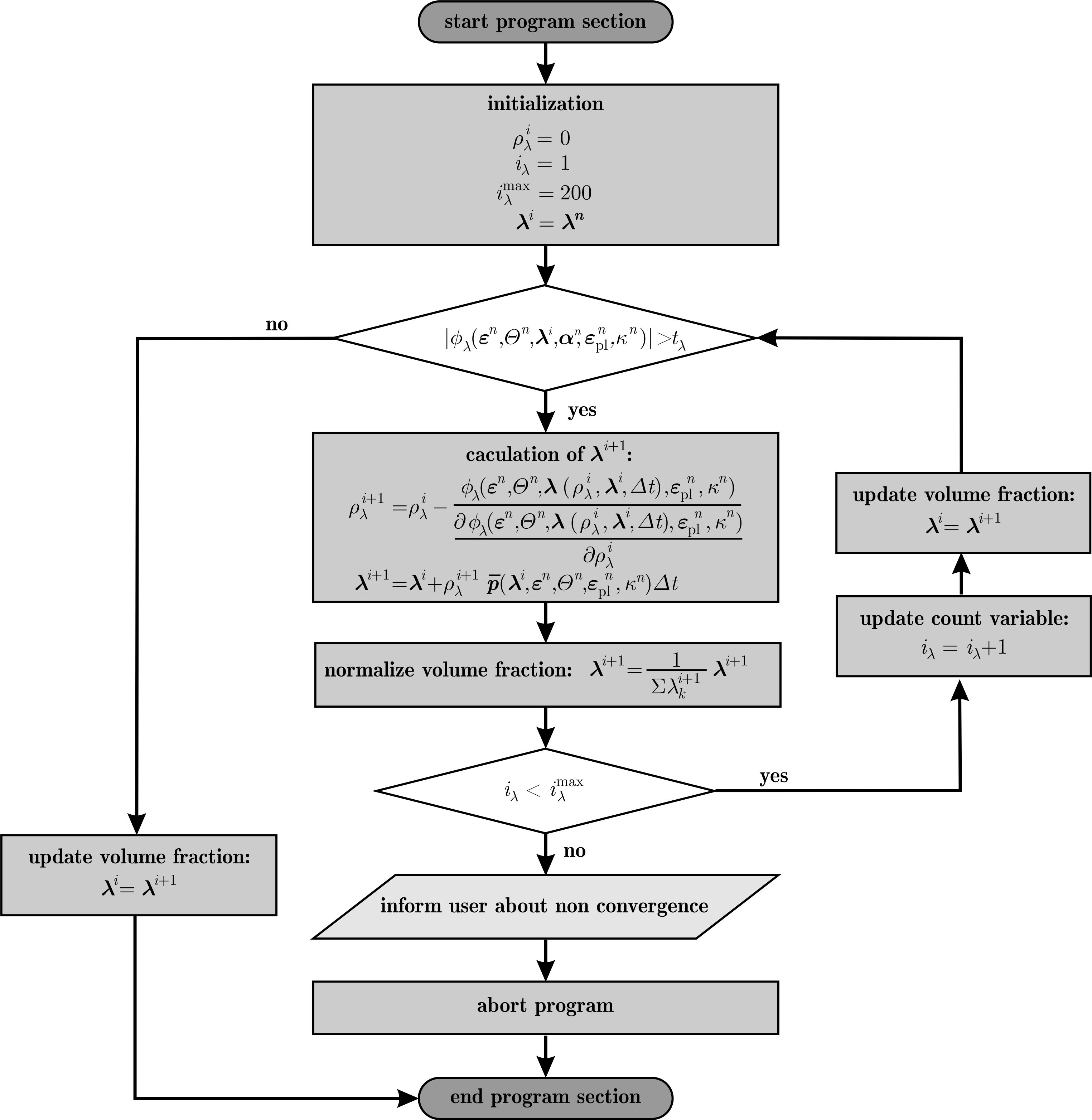}
\vspace{6pt}
\caption{Algorithm to calculate the volume fraction of the crystallographic phases $\pmb{\lambda}^{n+1}$}\label{fig:FlowChartLambda}
\end{figure}
\vspace{20mm}
\section{Numerical treatment in Abaqus}\label{sec4}
The UMAT subroutine in Abaqus must explicitly provide the Jacobian matrix of the constitutive model and the Cauchy stresses. The Cauchy stresses $\pmb{\sigma}^{n+1}$ can be determined using the general formulation for the stress rate
\begin{align}
\dot{\pmb{\sigma}} =
    \frac{\partial \, \pmb{\sigma}}{\partial \,\pmb{\varepsilon}}               \! \cdot \! \frac{\partial \, \pmb{\varepsilon}}{\partial \,t} 
    \, + \, \frac{\partial \, \pmb{\sigma}}{\partial \,\pmb{\lambda}}           \! \cdot \! \frac{\partial \, \pmb{\lambda}}{\partial \,t}
    \, + \, \frac{\partial \, \pmb{\sigma}}{\partial \,\pmb{\varepsilon}_{\mathrm{pl}}}  \! \cdot \! \frac{\partial \, \pmb{\varepsilon}_{\mathrm{pl}}}{\partial \,t}
    \, + \, \frac{\partial \, \pmb{\sigma}}{\partial \,\pmb{\alpha}}            \! \cdot \! \frac{\partial \, \pmb{\alpha}}{\partial \,t} \quad .
\end{align}
Discretizing this general formulation with the explicit Euler scheme results in 
\begin{align}
\pmb{\sigma}^{n+1} = \pmb{\sigma}^{n} 
    \,+ \, \frac{\partial \, \pmb{\sigma}}{\partial \,\pmb{\varepsilon}}        \bigg|^{n} \! \cdot \! \Delta\pmb{\varepsilon} 
    \, + \, \frac{\partial \, \pmb{\sigma}}{\partial \,\pmb{\lambda}}           \bigg|^{n} \! \cdot \! \Delta\pmb{\lambda} 
    \, + \, \frac{\partial \, \pmb{\sigma}}{\partial \,\pmb{\varepsilon}_{\mathrm{pl}}}  \bigg|^{n} \! \cdot \! \Delta\pmb{\varepsilon}_{\mathrm{pl}}
    \, + \, \frac{\partial \, \pmb{\sigma}}{\partial \,\pmb{\alpha}}            \bigg|^{n} \! \cdot \! \Delta\pmb{\alpha} \label{eq:stress}
\end{align}
where the partial derivatives are calculated as follows
\begin{align}
\frac{\partial \, \pmb{\sigma}}{\partial \,\pmb{\varepsilon}}\bigg|^{n} &= \overline{\mathbb{C}}^{n} \\
\frac{\partial \, \pmb{\sigma}}{\partial \,\lambda_{i}}\bigg|^{n} &=  - \Bigg[\Big( \pmb{Q}^{T} \cdot \eta_{i} \cdot \pmb{Q} + \overline{\mathbb{C}}_{i}^{-1} : \pmb{\sigma} \Big): \overline{\mathbb{C}} \Bigg]^{n} \\
\frac{\partial \, \pmb{\sigma}}{\partial \,\pmb{\varepsilon}_{pl}}\bigg|^{n} &= - \overline{\mathbb{C}}^{n}\\
\frac{\partial \, \pmb{\sigma}}{\partial \,\pmb{\alpha}}\bigg|^{n}&= 
\frac{\partial \, \pmb{\sigma}}{\partial \,\pmb{Q}}  :  \frac{\partial \, \pmb{Q}}{\partial \pmb{\alpha}} 
= \Bigg[ -\overline{\mathbb{C}}^{n} : \Bigg(  \frac{\partial \, \pmb{Q}^{T}}{\partial \pmb{\alpha}} \cdot \pmb{\overline{\eta}} \cdot \pmb{Q} + \pmb{Q}^{T} \cdot \pmb{\overline{\eta}} \cdot \frac{\partial \, \pmb{Q}}{\partial \pmb{\alpha}} \Bigg)\Bigg] :
\frac{\partial \, \pmb{Q}}{\partial \pmb{\alpha}} \quad . 
\end{align}
According to the Abaqus manual \cite{abaqusmanual} , the Jacobian matrix of the constitutive model is defined as \( \pmb{C}:=\frac{ \partial \, \Delta \pmb{\sigma}}{\partial \, \Delta \pmb{\varepsilon}} \). Considering this mathematical regularity, the Jacobian matrix is 
\begin{align}
\pmb{C}:= \frac{ \partial \, \Delta \pmb{\sigma}}{\partial \, \Delta \pmb{\varepsilon}} = \overline{\mathbb{C}}^{n} \quad .
\end{align}

\pgfplotsset{minor grid style={dotted,gray!70}}
\pgfplotsset{major grid style={dotted,gray!100}}
\pgfplotsset{
every axis legend/.style={
cells={anchor=center},
inner xsep=3pt,inner ysep=2pt,
nodes={inner sep=2pt,text depth=0.15em},
anchor=south east,
shape=rectangle,
fill=white,draw=black,
at={(0.98,0.02)},
},
}
\vspace{10mm}
\section{Numerical results}\label{sec6}
For all subsequent simulations, the material parameter set listed in Table \ref{tab1} is used. An essential part of the parameter set is taken from \cite{ junker2017numerical}. Herein, a brief discussion regarding the parameter identification of any given material is provided. Tensile testing enables the determination of the elastic constants \(E\) and \(\nu\) of the crystalline phases. Similarly, the scalar maximum transformation strain \(\hat{\eta}\) can be discerned using tensile testing, while an assumption is required for \(\hat{\nu}\) \cite{ junker2014accurate}. 
A temperature-dependent equation for the caloric part ${c}_{\mathrm{austenite}}$ of the Helmholtz free energy can be derived according to \cite{huo1993nonequilibrium}, utilizing measurement values from the DSC analysis. Alternatively, conducting tensile testing at different temperatures offers another method for establishing an equation for this caloric part. Given the nearly constant width of the hysteresis at different temperatures with only the plateau stresses changing, calibration of the material model based on caloric energy is feasible \cite{otsuka1999shape},\cite{bhattacharya2003microstructure}. By utilizing the two calibrated caloric energy values and their corresponding temperatures, a linear function with respect to temperature can be derived. 
As demonstrated in \cite{junker2016calibration}, this approach shows that accurate predictions of the mechanical material response for tensile tests at different temperatures are possible, even though the material parameters were derived from a thermal experiment, the DSC analysis. 
Consequently, it can be assumed that this linear function aptly characterizes the caloric part of the Helmholtz free energy, at least for simple boundary value problems. 

\noindent The dissipation parameter $r_{\lambda}$ can be calculated by utilizing measured values obtained from DSC analysis and subsequently evaluating an corresponding equation presented in \cite{JUNKER201586}. In contrast, determining of $r_{\alpha}$ still requires further research, and thus, a value for $r_{\alpha}$ must be assumed.
The parameter $\Lambda$ influences the magnitude of the introduced artificial energy. At \(\Lambda = 10^{-5} \), the magnitude of the artificial energy results in a remaining volume fraction of \(1 \%\) for the austenite phase at a uniaxial stress state and a strain of \(8 \%\). Any deviation exceeding this limit is considered to exceed the acceptable threshold of inaccuracy. Therefore, $\Lambda$ should not be chosen greater than $10^{-5}$ to ensure that its impact on the result remains minimal.
Nevertheless, the value should not be too low to provide sufficient robustness of the model.
\begin{table}[h]
\caption{Used material parameter set}\label{tab1}
\begin{tabular}{@{}lllllll@{}}
\toprule
parameter & value & unit &\textcolor{black}{\vline} & parameter & value & unit \\
\midrule
$E_{\mathrm{austenite}}$      & 83                       & [GPa]     &\textcolor{black}{\vline}& $k_{\text{0}}$              & 40                       & [GPa]\\
$E_{\mathrm{martensite}}$     & 40                       & [GPa]   &\textcolor{black}{\vline}& $k_{\text{1}}$              & 1                        & [GPa]\\
$\nu_{\mathrm{austenite}}$   & 0.35                     & [-]   &\textcolor{black}{\vline}& $k_{\text{2}}$              & 300                      & [-]\\
$\nu_{\mathrm{martensite}}$    & 0.35                     & [-]   &\textcolor{black}{\vline}& $\Lambda$                   & $5 \cdot 10^{-6}$        & [MPa]\\
$c_{\mathrm{austenite}}(\theta)$ & $-3.2465 - 0.51\theta$& [MPa]     &\textcolor{black}{\vline}&$r_{\alpha}$                & 1                        & [MPa]\\
$c_{\mathrm{martensite}}$     & 0                        & [MPa]   &\textcolor{black}{\vline}& $r_{\lambda}$               & 5.92                     & [MPa]\\
$\hat{\nu}$                 & 0.45                     & [-]   &\textcolor{black}{\vline}&$r_{\mathrm{pl}}$                    & 750                      & [MPa]\\
$\hat{\eta}$                & 0.055                    & [-]   &\textcolor{black}{\vline}& & & \\
\bottomrule
\end{tabular} 
\end{table}
\subsection{Investigation of the material model}\label{subsec1}
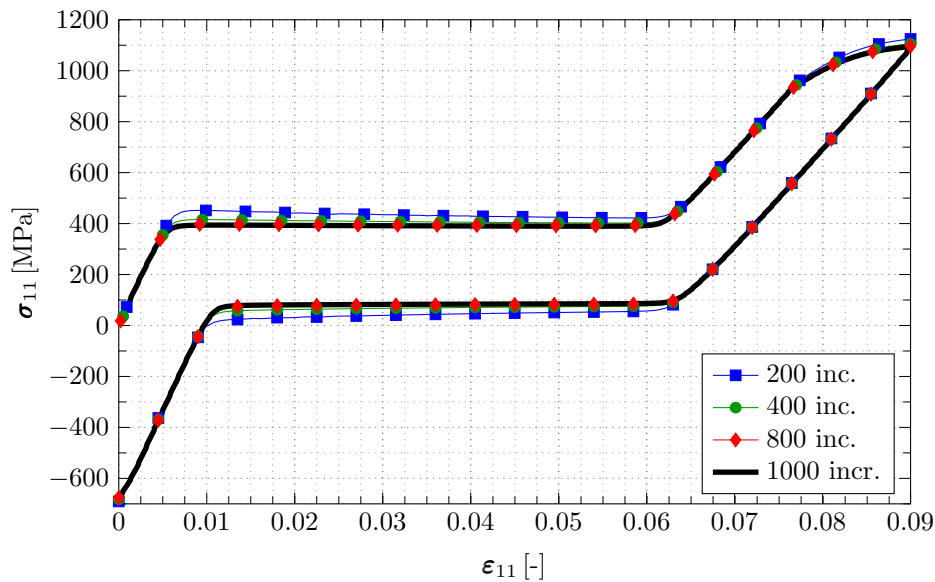
\begin{figure}[H]
    \centering
        \begin{tikzpicture}[scale=1.0]
            \begin{axis}[    
            xlabel={$\pmb{\varepsilon}_{11}\,$[-]},
            ylabel={$\pmb{\sigma}_{11}\,$[MPa]},
            xmin=0, xmax=0.09,
            ymin=-700, ymax=1200,
            xtick={0,0.01,...,0.09},
            ytick={-600,-400,...,1200},
            minor xtick={0.0025,0.005,0.0075,0.0125,0.015,...,0.09},
            minor ytick={-700,-500,-300,-100,100,...,1100},
            log ticks with fixed point,	
            ticklabel style={
            	/pgf/number format/.cd,
            	/pgf/number format/fixed,
            	1000 sep = {}
            },
            legend style={
            legend cell align=left 
            },
            scaled ticks=false,
            grid=both,
            axis line style={black},
            tick style={black},
            width=12cm,
            height=8cm
            ]
            \addplot[color=blue,mark=square*,mark options={scale=1.0,mark repeat=5,mark phase=2}] table[x index=0,restrict x to domain=0.0:0.09, y index=1, col sep=space]{200Inkr.dat};
            \addplot[color=green!60!black,mark=*,mark options={scale=1.0,mark repeat=10,mark phase=2}] table[x index=0, y index=1, col sep=space]{400Inkr.dat};
            \addplot[color=red,mark=diamond*,mark options={scale=1.3,mark repeat=20,mark phase=2}] table[x index=0, y index=1, col sep=space]{800Inkr.dat};
            \addplot[color=black, line width=2pt,mark=none] table[x index=0, y index=1, col sep=space]{1000Inkr.dat};             
            \legend{
                {200 inc.},
                {400 inc.},
                {800 inc.},
                {1000 incr.}
            }    
            \end{axis}
        \end{tikzpicture}
    \caption{Stress-strain behaviour depending on the total number of load increments}\centering
    \label{fig:TensionTest}
\end{figure}

Figure \ref{fig:TensionTest} illustrates the material response during a tensile test as a function of the total number of loading steps. The material response refers to a hexahedron element of type C3D8R. The boundary conditions were selected so that a uniaxial stress state is present in the entire element.
All curves shown qualitatively exhibit the same material behavior characteristic of the superelasticity effect. 
Initially, a linear increase is observable, with the slope corresponding to the Young's modulus of the austenite phase. The transition from linear behavior to a plateau signifies the beginning of the phase transformation from austenite to martensite. The completion of the phase transformation is indicated by the transition from plateau to a linear slope.
The slope after the end of the phase transformation corresponds to the Young's modulus of martensite and is therefore different from the slope at the beginning of the loading.  Once the yield point is exceeded, nonlinear hardening occurs.
After applying 9$\%$ strain, stepwise unloading begins. At a certain point, the reverse transformation occurs until the plateau returns to a linear progression. The Element is deformed back to its original shape despite existing plastic deformation, leading to the occurrence of compressive stresses at the end of the loading.
\noindent It can be observed that as the number of loading steps increases, the behavior approaches that of the black curve. A reasonable accuracy is achieved with a number of 400 increments. Importantly, it is not necessary to define at least 400 load increments for every boundary value problem. The tensile test in Fig. \ref{fig:TensionTest} represents an extreme case, since a complete phase transformation of one phase into the other phase occurs twice. For boundary value problems where the extent of phase transformation is much smaller, the number of load increments can be significantly reduced without affecting the accuracy of the solution. 
Furthermore, it should be noted that the dependence of the material behavior on the number of load increments is due to the explicit time integration method used and not to the evolution equations themselves.

\begin{figure}[H]%
\centering
\includegraphics[width=0.9\textwidth]{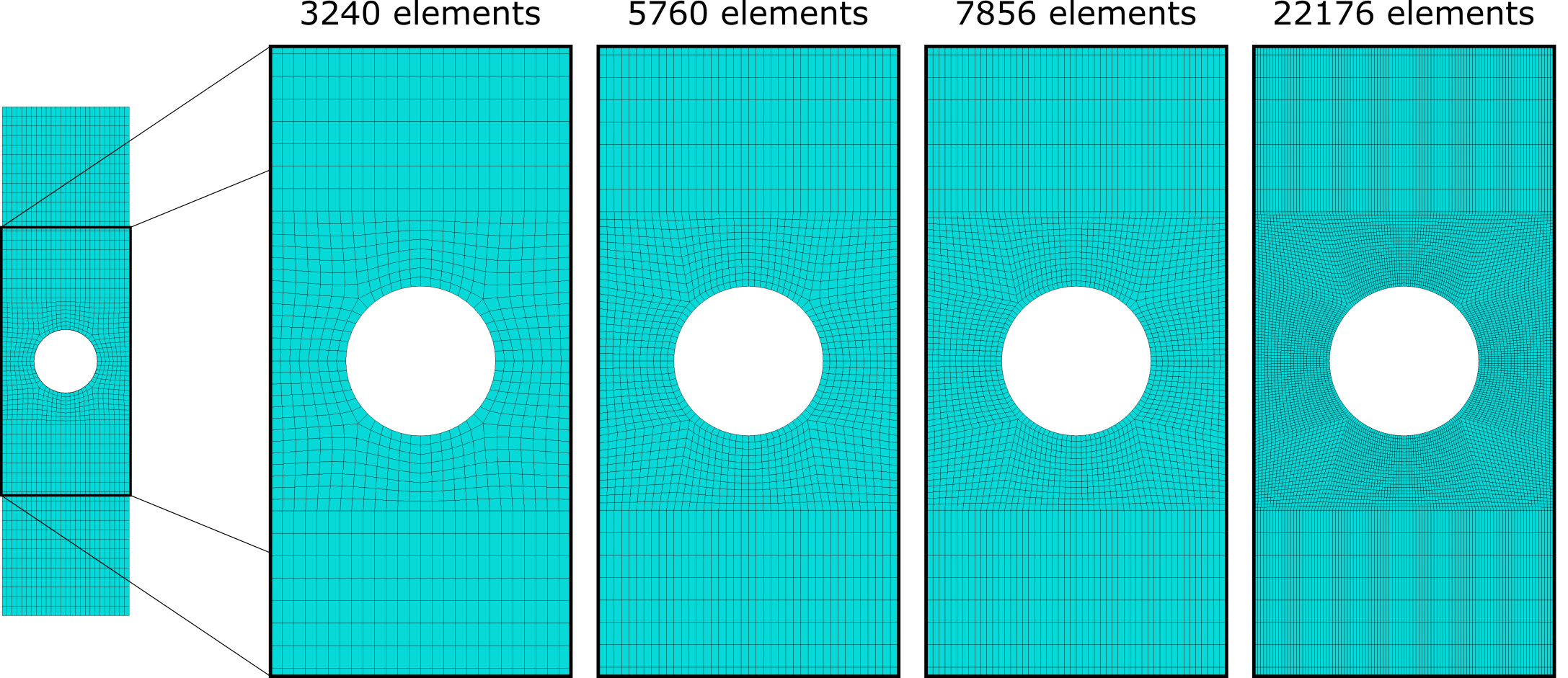}
\caption{Different mesh finenesses with the C3D8R hexahedron element to investigate mesh dependency}\label{fig:Mesh_fineness}
\end{figure}

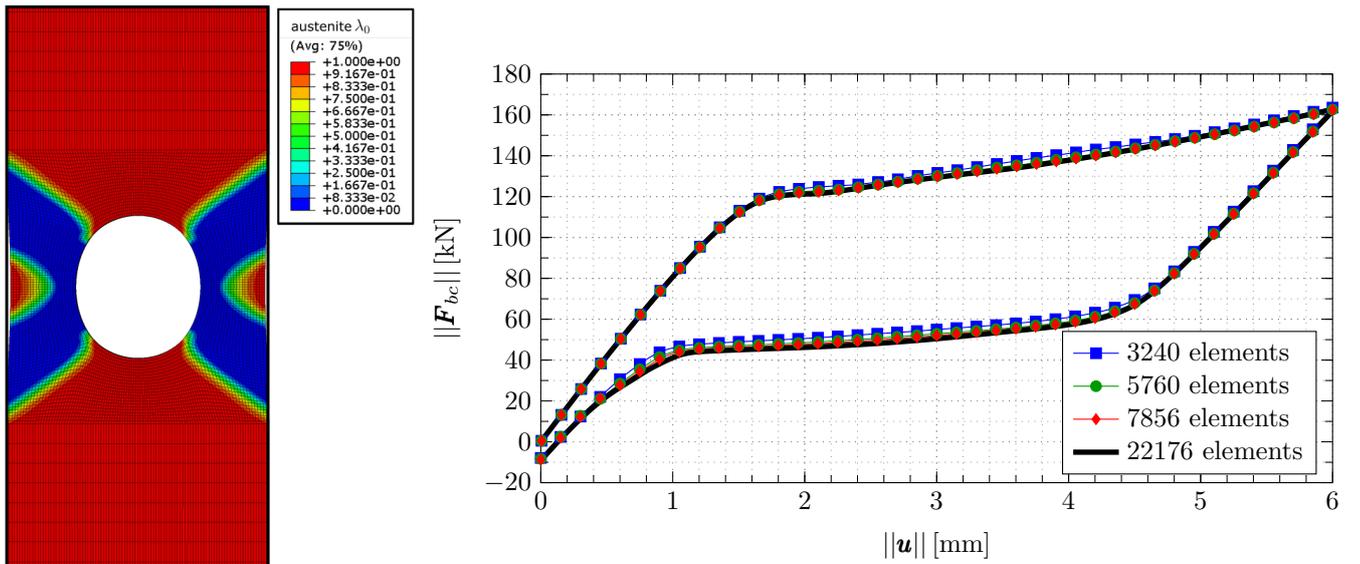
\begin{figure}[H]
        \begin{subfigure}[b]{0.32\textwidth}
        \includegraphics[width=\linewidth]{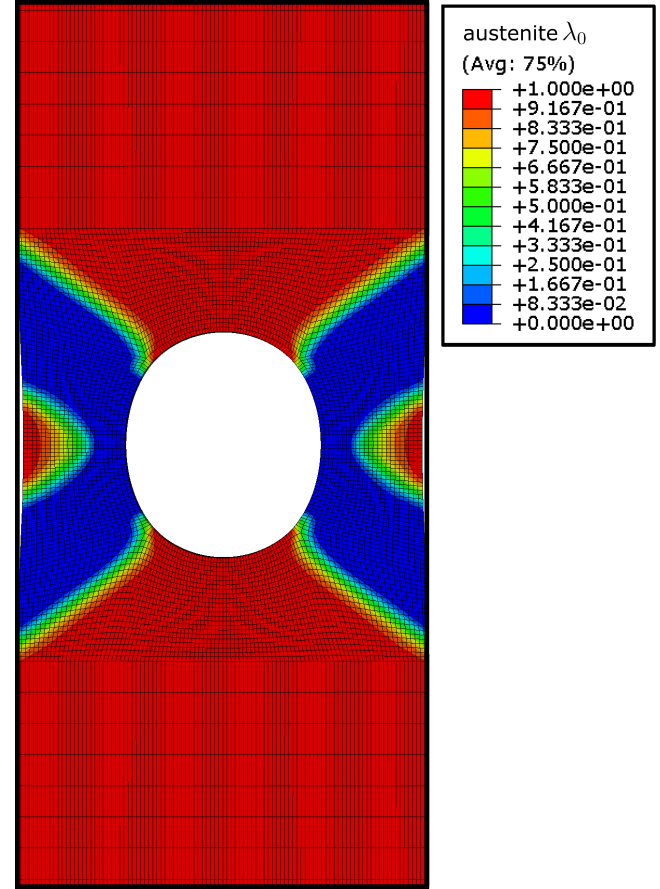} 
        \end{subfigure}
    \begin{subfigure}[b]{0.3\textwidth}
        \begin{tikzpicture}[scale=1.0]
            \begin{axis}[    
            xlabel={$||\pmb{u}||\,$[mm]},
            ylabel={$||\pmb{F}_{bc}||\,$[kN]},
            xmin=0, xmax=6,
            ymin=-20, ymax=180,
            xtick={0,1,...,6},
            ytick={-20,0,...,180},
            minor xtick={0.2,0.4,0.6,0.8,1.2,1.4,...,5.8},
            minor ytick={-10,10,30,40,50,...,170},
            log ticks with fixed point,	
            ticklabel style={
            	/pgf/number format/.cd,
            	/pgf/number format/fixed,
            	1000 sep = {}
            },
            legend style={
            legend cell align=left 
            },
            scaled ticks=false,
            grid=both,
            axis line style={black},
            tick style={black},
            width=12cm,
            height=7cm
            ]
            \addplot[color=blue,mark=square*,mark options={scale=1.0,mark repeat=25,mark phase=2}] table[x index=0, y index=1, col sep=space]{3240elements.dat};
            \addplot[color=green!60!black,mark=*,mark options={scale=1.0,mark repeat=25,mark phase=2}] table[x index=0, y index=1, col sep=space]{5760elements.dat};
            \addplot[color=red,mark=diamond*,mark options={scale=1.0,mark repeat=25,mark phase=2}] table[x index=0, y index=1, col sep=space]{7856elements.dat};
            \addplot[color=black, line width=2pt,mark=none] table[x index=0, y index=1, col sep=space]{22176elements.dat};  
            \legend{
                {3240 elements},
                {5760 elements},
                {7856 elements},
                {22176 elements}
            }    
            \end{axis}
        \end{tikzpicture}
    \end{subfigure}
    \caption{Volume fraction distribution of the austenite phase in the FE mesh with 22179 elements at maximum load and reaction force-displacement curves for different degrees of mesh fineness}
    \label{fig:MeshConvergence_loadDisplacement}
\end{figure}

\noindent Having focused on the results and observations of the tensile test, it is also important to investigate the reliability and accuracy of the material model. To ensure the independence of our simulation results from the mesh fineness and thus confirm their reliability, we performed a mesh convergence study on a plate measuring 400mm by 100mm by 5mm, featuring a hole with a diameter of 50mm. The different degrees of fineness of the FE mesh is shown in Fig. \ref{fig:Mesh_fineness}.
\noindent The plate is loaded with a displacement-based tensile load and then returned to its original position. The corresponding reaction force is plotted against the displacement-based load, as shown in Fig. \ref{fig:MeshConvergence_loadDisplacement}. Initially, the austenite phase is present due to the prevailing temperature in the plate. As the load rises, the volume fraction of martensite in the plate continues to increase. As expected, this transformation is reflected in a degressive increase of the force in the force-displacement diagram. When the maximum displacement-based tensile load of 6 mm is reached, a progressive unloading phase begins. In this phase, a reversal of the transformation from martensite back to austenite occurs, resulting in a hysteresis. After loading and unloading, a negative reaction force is observed at the 0-mm position, indicating a compressive force. This observation proves that plastic strain must be present in the notches after the loading phase.
When looking at the material response for the different meshes, it is noticeable that the responses are relatively close, even though the coarsest mesh has about 7 times fewer elements. Nevertheless, with increasing mesh fineness, a tendency can be observed with regard to the material behavior. As the mesh size increases, the material response approaches the black curve associated with the finest mesh. From this, it can be deduced that the material model is mesh independent.
\subsection{Stent under processing and operating conditions}\label{subsec4}
\begin{figure}[H]%
\centering
\includegraphics[width=0.7\textwidth]{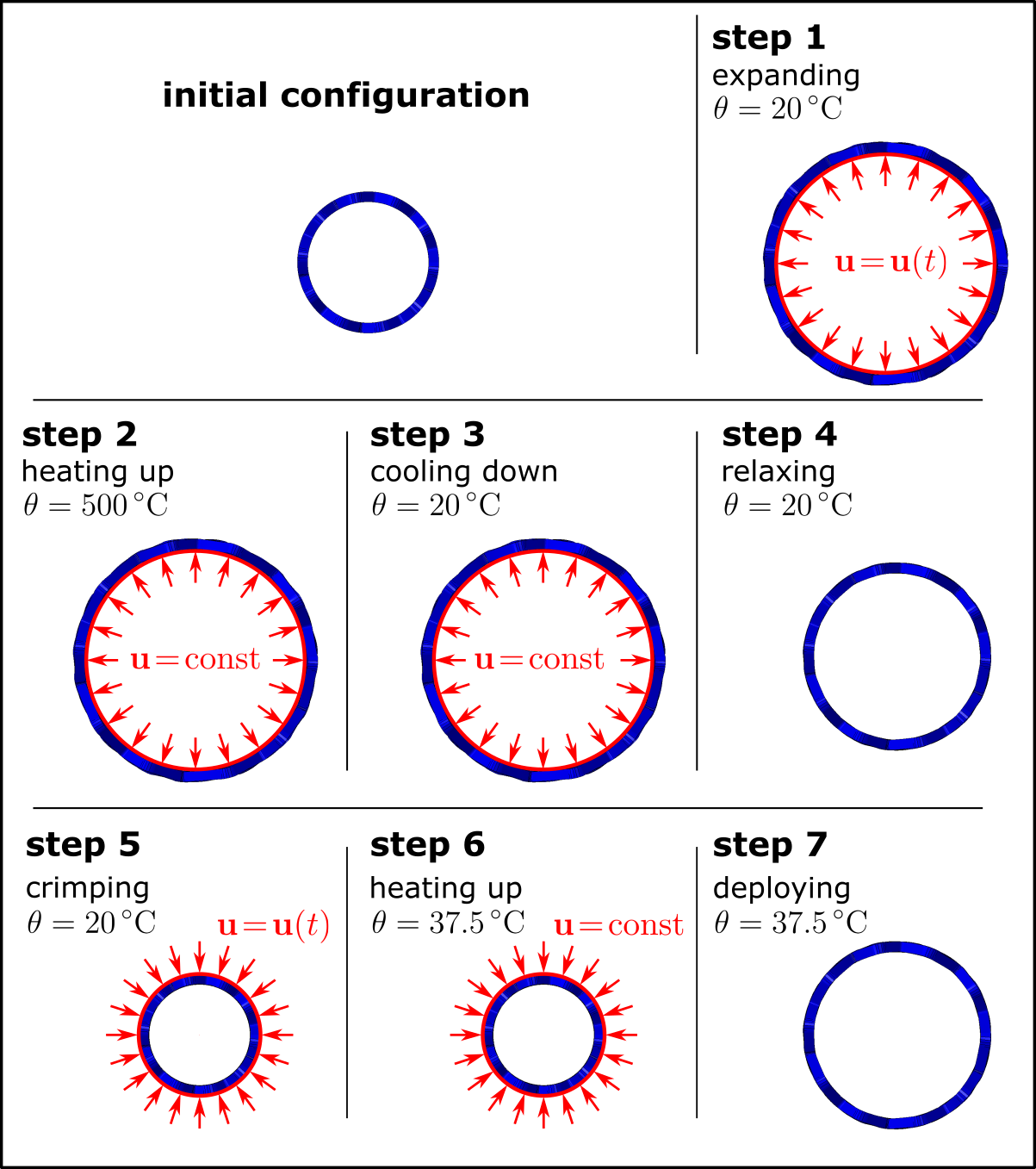}
\caption{Defined boundary conditions in each time step in stent boundary value problem}\label{fig:stent_boundaryCondition}
\end{figure}

\begin{figure}[H]%
\centering
\includegraphics[width=1.0\textwidth]{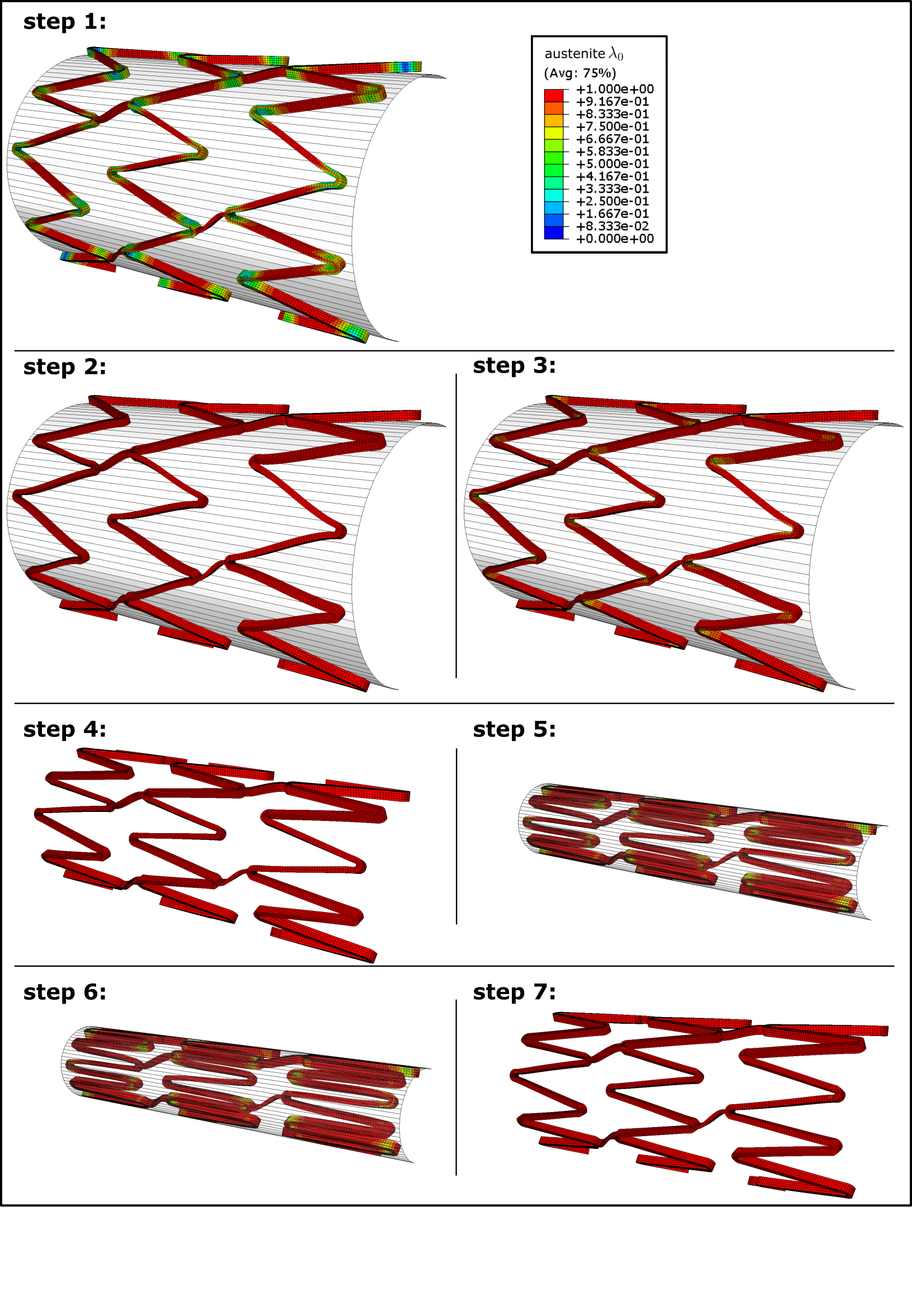}
\vspace{-20mm}
\caption{Austenite volume fraction after each process step within the stent shown in the half-section}\label{fig:VolumeFractionOverAllTimeSteps}
\end{figure}

\begin{figure}[H]%
\centering
\includegraphics[width=1.0\textwidth]{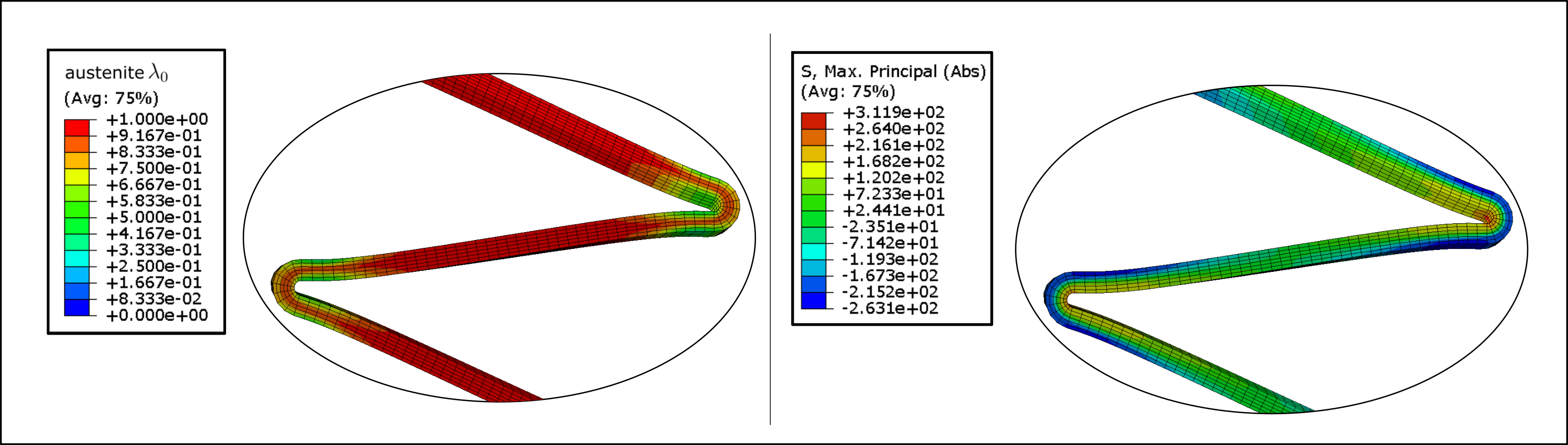}
\caption{Comparison between the austenite volume fraction and the maximum principal stress within a region in the stent after expansion}\label{fig:comparism_VolumeFractionStress}
\end{figure}

\begin{figure}[H]
    \centering
        \begin{tikzpicture}[scale=1.0]
            \begin{axis}[    
            xlabel={$\text{t}\,$[s]},
            ylabel={$\text{u}_\text{radial}\,$[mm]},
            xmin=0, xmax=7,
            ymin=-2, ymax=5,
            xtick={0,1,...,7},
            ytick={-1,0,1,...,5},
            minor xtick={0.2,0.4,0.6,0.8,1.2,1.4,...,6.8},
            minor ytick={-2,-1.8,-1.6,-1.4,-1.2,-0.8,-0.6,...,5},
            log ticks with fixed point,	
            ticklabel style={
            	/pgf/number format/.cd,
            	/pgf/number format/fixed,
            	1000 sep = {}
            },
            legend style={
            legend cell align=left 
            },
            scaled ticks=false,
            grid=both,
            axis line style={black},
            tick style={black},
            width=12cm,
            height=6cm
            ]
             \addplot[color=blue, line width=2pt,mark=none] table[x index=0, y index=1, col sep=space]{RadialDisplacement_Time.dat};    
            \end{axis}
        \end{tikzpicture}
    \caption{Radial displacement of a arbitrary node  of the stent as a function of the time}\centering
    \label{fig:RadialDisplacementArbitraryNode}
\end{figure}
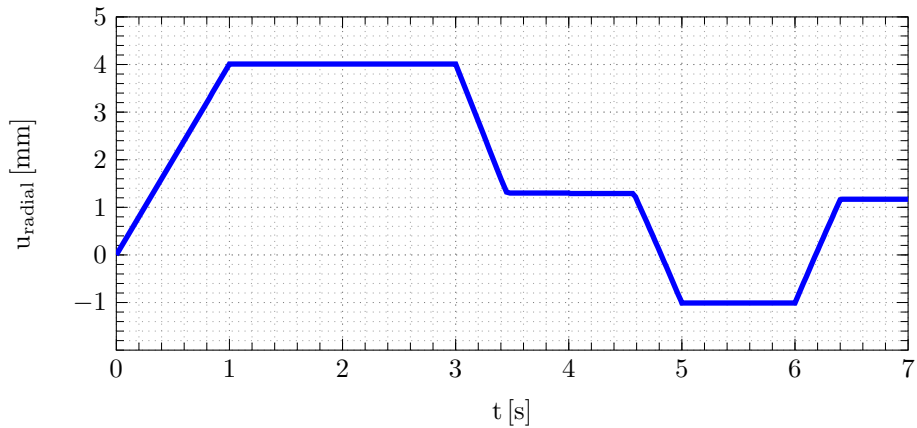

In this section, we consider a boundary value problem in which a stent is subjected to processing and operating conditions. The defined boundary conditions in each calculation step can be seen in Fig. \ref{fig:stent_boundaryCondition}. In the first step, radial displacement is applied to a rigid surface at the inner radius of the stent, resulting in stent expansion. Then, in steps 2 and 3, heat treatment is performed while the stent is in the expanded state by heating the stent up to 500$^\circ \mathrm{C}$ and cooling it down to 20$^\circ \mathrm{C}$. 
\noindent In step 4, the applied radial displacement is gradually removed, resulting in some degree of elastic recovery. Steps 1 to 4 describe the shaping process.
In step 5, radial displacement is applied to a second rigid surface that grips the stent at its outer surface, resulting in crimping. The stent is then heated to body temperature while in this locked state.
In the final step, the boundary condition of displacement of the second rigid surface is gradually removed, allowing the stent to deploy.
\noindent 
Having defined the boundary value problem, its simulation results are examined. For this purpose the austenite volume fraction after each process step is shown in Fig. \ref{fig:VolumeFractionOverAllTimeSteps}. At the beginning, only the austenite phase is present. After the first process step, the martensite phase has evolved as expected in highly stressed areas. This phenomenon is referred to as stress-induced martensite. The temperature increase from 20°C to 500°C causes the martensite areas to transform back again, as the austenite phase is energetically more favorable. When the temperature of the stent is lowered back to 20°C in step 3, there is less martensite than after process step 1 due to the dissipative characteristic. A high austenite content is necessary to ensure the pseudoelasticity effect in the subsequent process. As the load is removed in step 4, the stress-induced martensite is completely transformed back to austenite. Crimping takes place in step 5. As a result, the martensite phase reappears in highly stressed areas and remains despite the increase to body temperature in step 6. In step 7, the radial load is removed. As a result, the stent deploys. This phenomenon, known as the pseudoelasticity effect, is based on the phase transformation of the martensite phase back into the austenite phase.
\noindent To examine the effect of stress-induced martensite in more detail, the comparison between the maximum principal stress and the volume fraction of austenite phase is shown graphically in Fig. \ref{fig:comparism_VolumeFractionStress}. 
The comparison reveals that highly stressed areas characterized by high maximum principal stress exhibit a lower volume fraction of austenite. It can be deduced that as the maximum principal stress increases, the volume fraction of martensite increases accordingly. This correlation is consistent with the well-known phenomenon in the literature known as stress-induced martensite, as shown in \cite{otsuka1999shape}. It is also visible that a neutral fiber is present due to the bending load. 
As expected, only the austenite phase is present in this neutral fiber.
\vspace{-5mm}
\noindent 
In Fig. \ref{fig:RadialDisplacementArbitraryNode}, the radial displacement of a representative node from the surface of the stent is plotted over time.
The choice of this specific node is of little importance, as almost all nodes have the same radial displacement. The diagram illustrates an increase in the stent's radius by approximately 1.3 mm following the forming process. 
 The diagram also shows that the stent does not return to its original configuration after crimping. Additionally, it is evident from the diagram that the stent does not return to its original shape after crimping. This indicates that a small amount of the martensite phase is present within the stent after the forming process. 
\newpage
\subsection{Compression staple under processing and operating conditions}\label{subsec5}
\begin{figure}[H]%
    \centering
    \includegraphics[width=0.5\textwidth]{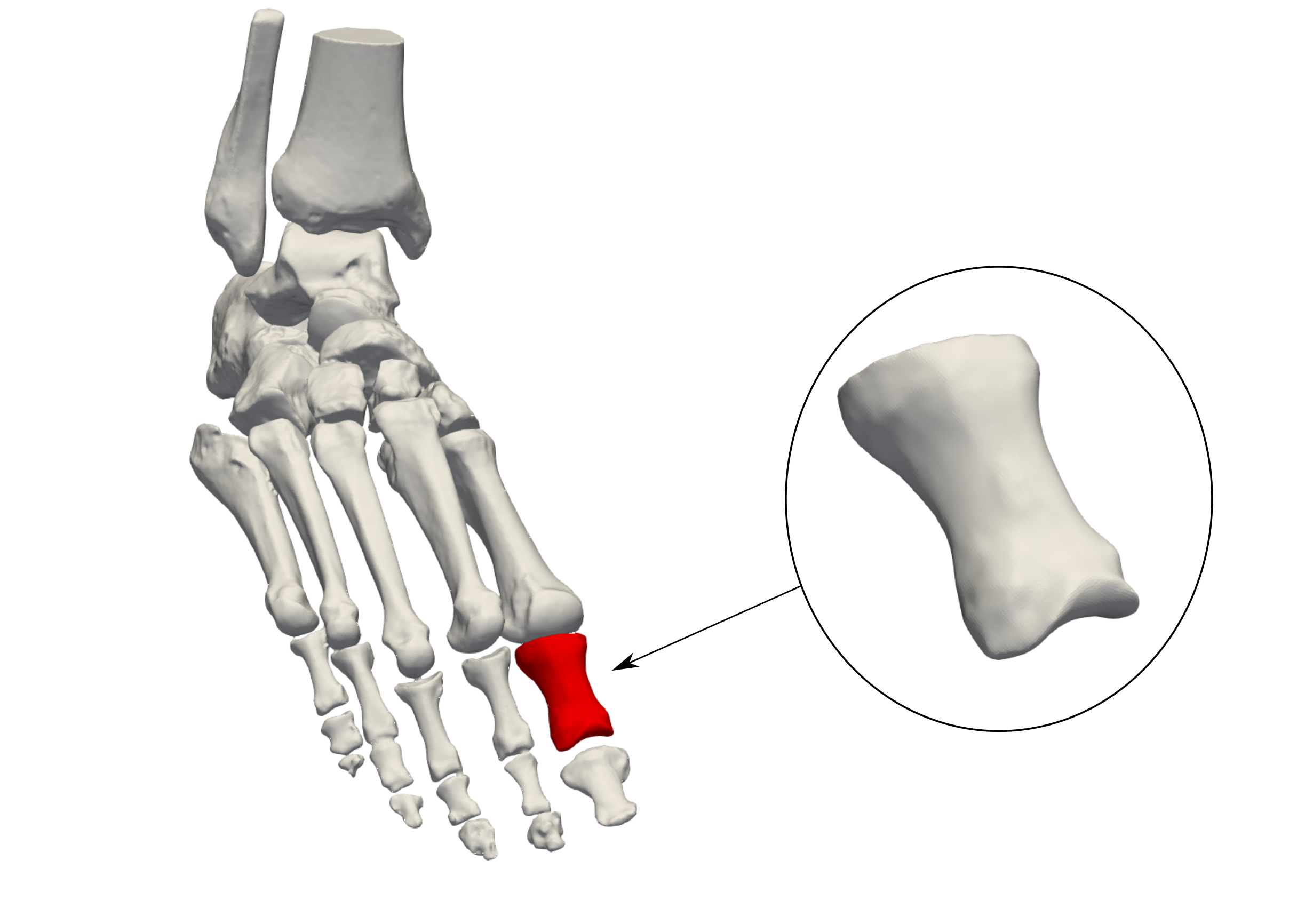}
    \caption{3D model of a human foot \cite{footbonemodel}
 }\label{fig:ToeBone} 
\end{figure}

\begin{figure}[H]%
\centering
\includegraphics[width=0.67\textwidth]{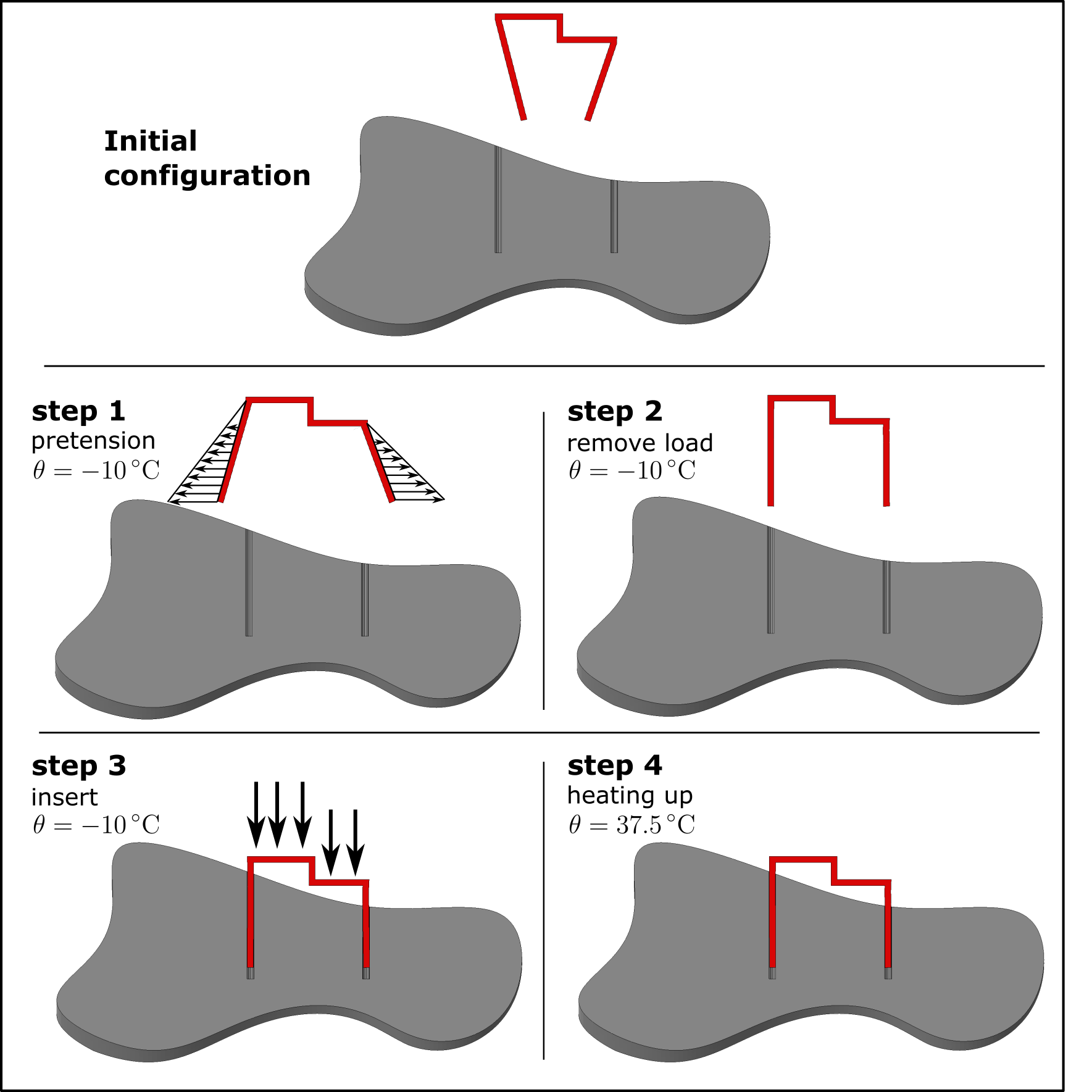}
\vspace{10pt} 
\caption{Schematic illustration of defined boundary conditions in each time step for the staple boundary value problem}\label{fig:PrincipleProcedure_staplesimulation}
\end{figure}

\begin{figure}[H]
    \centering
        \begin{tikzpicture}[scale=0.7]
            \begin{axis}[    
            xlabel={$\pmb{\theta}$ [°C]},
            ylabel={$\pmb{F}_{c}$ [N]},
            xmin=-10, xmax=40,
            ymin=0, ymax=5,
            xtick={-10,-5.0,0,...,40},
            ytick={0,1,...,5},
            minor xtick={-8.75,-7.5,-6.25,-3.75,-2.5,-1.25,...,40},
            minor ytick={0,0.2,0.4,0.6,0.8,1.2,1.4,...,5},
            log ticks with fixed point,	
            ticklabel style={
            	/pgf/number format/.cd,
            	/pgf/number format/fixed,
            	1000 sep = {}
            },
            legend style={
            legend cell align=left 
            },
            scaled ticks=false,
            grid=both,
            axis line style={black},
            tick style={black},
            width=15cm,
            height=12cm
            ]
            \addplot[color=blue,mark=*,mark options={scale=0.5}] table[x index=0,y index=1, col sep=space]{temperature_force.dat};
            \legend{
            }    
            \end{axis}
        \end{tikzpicture}
    \caption{Normal Contact force between the leg of the staple and the drilled hole depending on the temperature in step 4}\centering
    \label{fig:compressionTemperature}
\end{figure}
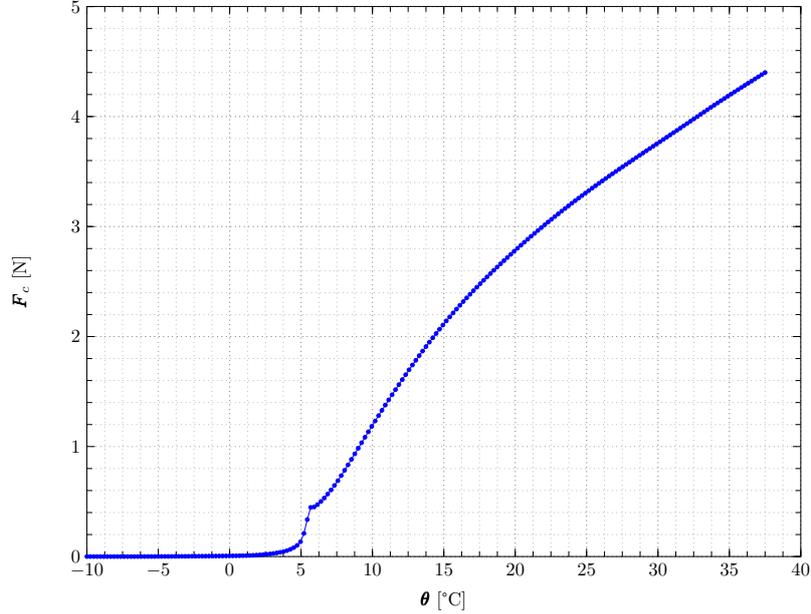

\begin{figure}[H]%
\centering
\includegraphics[width=1.0\textwidth]{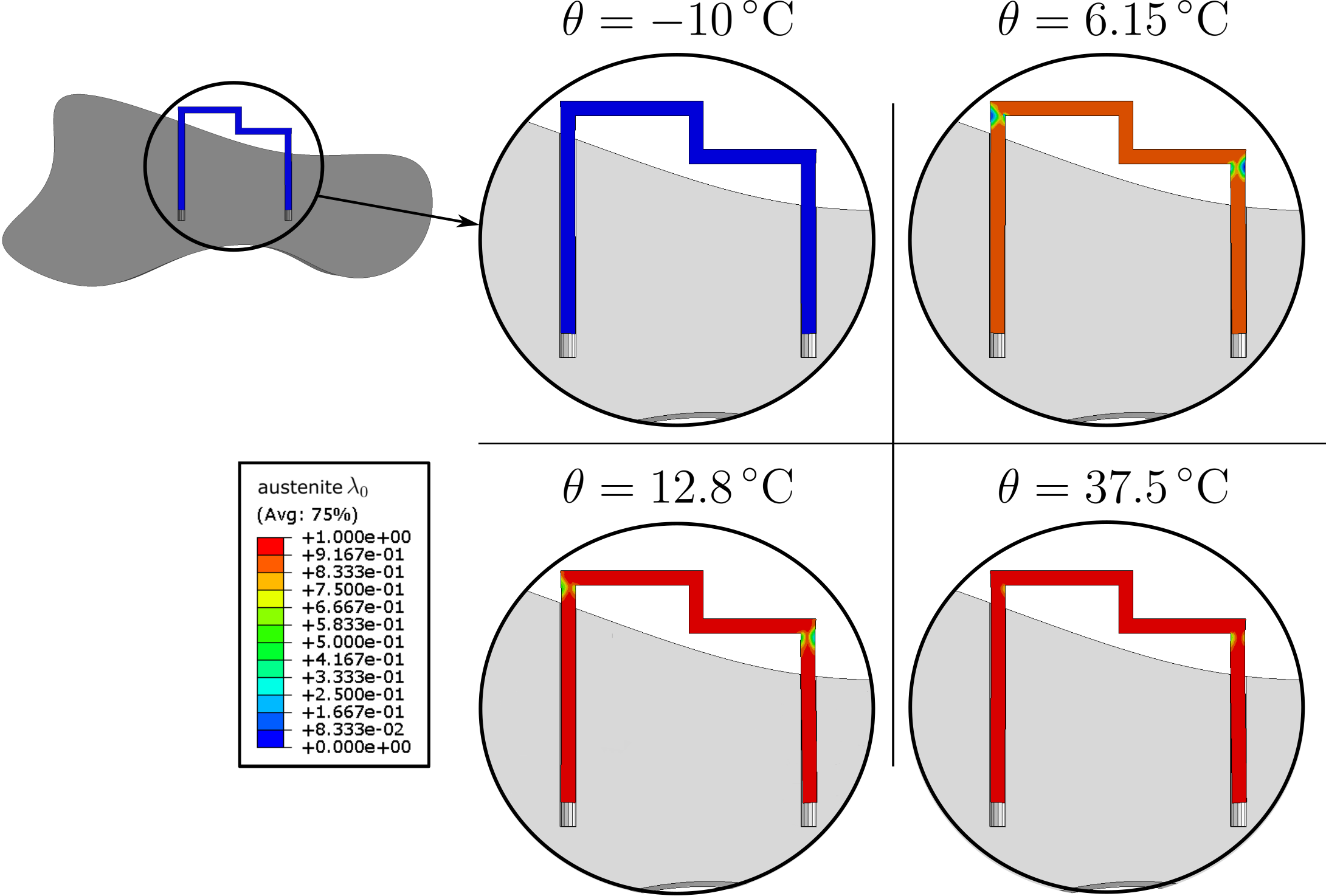}
\caption{Distribution of the austenite volume fraction in the staple with hidden mesh}\label{fig:ToeBone_volumefraction}
\end{figure}

Compression staples made of shape memory alloys are used in the treatment of bone fractures since they can improve and accelerate the healing process \cite{mereau2006nitinol}. Activated by the body's temperature or other techniques, these staples exhibit the shape memory effect, leading them to close and apply compression to the fracture gap. The staples are often used for fractures of the so-called phalanges bones, which are located in the hand and foot. The bone of the modelled foot marked in red in the Fig. \ref{fig:ToeBone} is one of the phalanges bones. 
\vspace{-18pt} 
\vspace{-20pt} 
\noindent In this boundary value problem, an abstract geometry of this red marked phalange bone was used. For the bone, a linear-elastic behavior with isotropic elasticity is assumed, which seems to be a reasonable approximation, especially considering the small forces involved. The elastic constants for the bones come from \cite{qiu2011finite} and are $E_{\text{bone}}=7300$ and $\nu_{\text{bone}}=0.3$, respectively. 
The defined boundary conditions are based on the conventional method described by \cite{mereau2006nitinol} for the processing and application of these staples made of shape memory alloys. In Fig. \ref{fig:PrincipleProcedure_staplesimulation}, the boundary conditions of the boundary value problem are illustrated in each process step. 
\vspace{-12pt} 
\vspace{-18pt} 
\noindent 
Initially, the compression staple is subjected to a displacement-based load on its legs in the first step. Subsequently, in the second step, the displacement-based load is eliminated, resulting in elastic re-deformation of the staple. The following step involves inserting the staple into the drilled hole in the bone. Finally, by increasing the temperature in the last step, the shape-memory effect is activated, resulting in compression within the region situated between the two holes. The extent of compression is defined as the resulting contact force between the leg of the staple and the bore hole. This quantity is shown in the last process step as a function of temperature in Fig. \ref{fig:compressionTemperature}. 
\noindent The diagram illustrates that there is no contact force between the clamp and the bone up to approximately 5$^\circ$C, meaning the clamp does not exert any compressive force on the bone. It is only from about 5$^\circ$C that pressure exerted on the drill hole, with the compressive force increasing as the temperature rises.
As anticipated, Fig. \ref{fig:ToeBone_volumefraction} illustrates that the increase in contact force is attributed to a crystallographically reversible lattice transformation from the martensite phase to the austenite phase, known as the shape memory effect.

\noindent At the beginning of time step 4, only the martensite phase is visually discernible.
After heating to 6.15$^\circ$C, a distinctly different phase composition becomes apparent.
The volume fraction of the austenite phase predominates in the large part of the clamp. 
Only the areas heavily stressed in the 'pretension' time step exhibit a higher volume fraction of the martensite phase.
This phenomenon is consistent with expectations, as residual strains are present in these regions after loading in step 1 and unloading in step 2, which increase the stability of the martensite phase. These regions require higher temperatures to trigger the transformation of the martensite phase into the austenite phase.
Thus, an increase in temperature from 12.8$^\circ$C to body temperature of 37.5$^\circ$C results in a more extensive volume fraction of austenite.
It is worth noting that, in general, a microstructure of pure austenite or pure twinned martensite should macroscopically result in identical shapes, which is known as the one-way effect. In this boundary value problem, uniformly distributed martensite, representing twinned martensite, is present in the initial state.
Consequently, the tendency to return to the initial state should grow with increasing volume fraction of austenite in the final step.
This physical phenomenon can be represented by the material model, as corroborated by the simulation results. 
The tendency to return to the initial state increases with rising temperature, which is reflected in the increasing contact force.
 
\newpage
\section{Conclusions}\label{sec7}
In this paper, a material model for shape memory alloys (SMA) has been presented which is characterized by its robustness and provides reasonable results. These insights are derived from the evaluation of results of two complex boundary value problems from the field of medical technology. The first boundary value problem relates to the simulation of a stent, undergoing the entire process chain from the manufacturing process to the application process. 
This boundary value problem is deemed highly complex, due to existing contact conditions and significant temperature changes under intense mechanical stresses. The second boundary value problem involves the simulation of a clamp for healing a fracture in the foot bone, where the mechanism of the one-way effect under contact conditions also presents a complex boundary value problem. 
Further, the mesh convergence study revealed that the material model is mesh-independent.

\noindent The characteristic capabilities of the material model compared to other models result from several implemented modifications. One of these modifications is the use of rate-independent evolution equations for internal variables describing the phase transformation. A rate-independent evolution, corresponds to the real physical processes in shape memory alloys, since phase transformations are diffusionless, which is equivalent to a time-independence or rate-independence.
The use of rate-independent evolution equations is facilitated by employing effective algorithms, which are presented in the third section.
These algorithms subsequently correct an error arising from the explicit time integration of the evolution equations, leading to an inaccurate determination of the Lagrange multipliers. They provide a simple feasible alternative compared to the complex exponential mapping time integration method of \cite{weber1990finite}.
By correcting this error, the algorithms contribute to the enhancement of the model's robustness.
An additional improvement includes incorporating a penalty term and substituting the volume fractions with a sigmoid function. This means that it is no longer necessary to use the active set method. The active set method is known to cause issues with convergence, particularly when there are frequent changes in the active set.
Another modification enhancing the capabilities of the material model is the parametrization of the rotation matrix using Euler-Rodrigues parameters instead of Euler angles. With Euler-Rodrigues parameters, there is no risk of Gimbal Lock, as is the case with Euler angles.
The last modification concerns the representation of twinned martensite. By a suitable choice of initial Euler-Rodrigues parameters, the material behavior does not depend on the loading direction, a characteristic of twinned martensite. By selecting initial Euler-Rodrigues parameters based on a certain principle, unfavorable initial angles that negatively impact the model’s convergence behavior and stability are avoided.
 
\noindent Considering the aforementioned aspects, the model makes an important contribution to understanding the complex thermomechanical properties of SMA.
Due to its applicability it is a valuable tool for various industrial sectors, including medical technology, automotive industry, aerospace, and robotics, to explore and develop innovative applications of SMA. The model's practical orientation allows for a realistic simulation of SMA components and thus, addresses the industrial demand for reliable and efficient methods for material modeling. The exemplary applications, like stents and clamps, underline the model's relevance and versatility in various application areas.
  
\noindent Despite its advantages and comprehensive applicability, some possibilities for further development and refinement of the model remain open. 
For example, future work may address the extension of the material model regarding large deformations. Another development possibility may consider the inclusion of dynamic effects. This advancement will open new possibilities for precisely predicting the behavior of SMA.
\vspace{3mm}\\
\textbf{Acknowledgements}
CE and PJ were funded by the Deutsche Forschungsgemeinschaft (DFG, German Research Foundation) through the project grant TRR 298 (SIIRI), for which we are profoundly thankful. TB and PJ are grateful for the support provided by the Deutsche Forschungsgemeinschaft (DFG) under Germany’s Excellence Strategy within the cluster of Excellence PhoenixD (EXC 2122, Project ID 390833453). 
The authors furthermore thank V. Meine for his support in the creation of the plots.

\newpage
\begin{appendices}

\section{Figures}\label{secA1}

\begin{figure}[H]%
\centering
\includegraphics[width=0.35\textwidth]{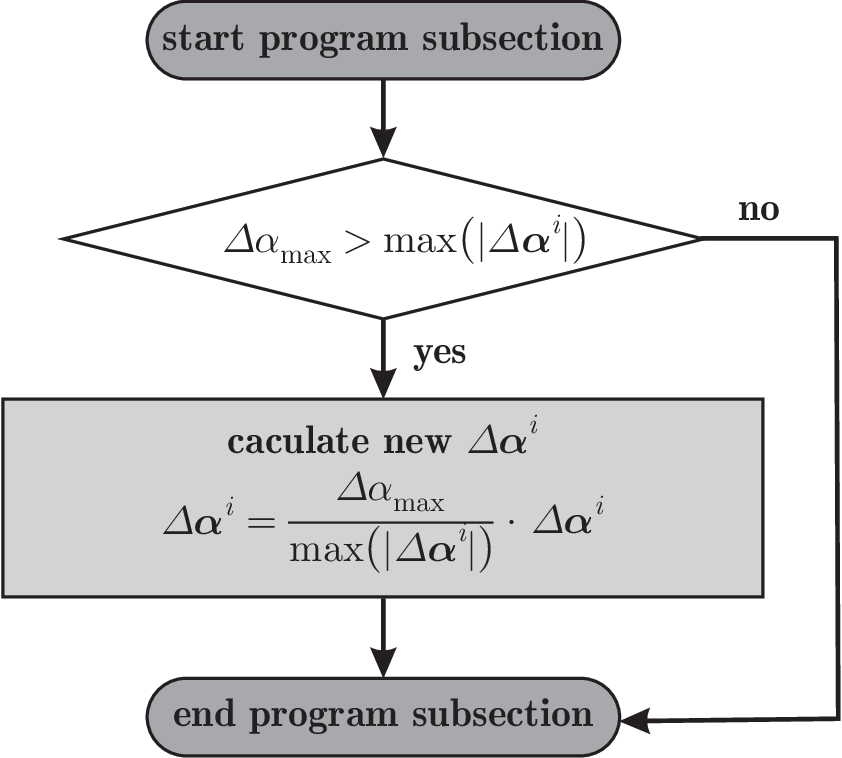}
\vspace{3mm}
\caption{Subsection of user subroutine UMAT to limit $\Delta \pmb{\alpha}^{i}$ when calculating Euler-Rodrigues parameter}\label{fig:ProgramSubsectionLimitDeltaAlpha}
\end{figure}
\begin{figure}[H]%
\centering
\includegraphics[width=1.0\textwidth]{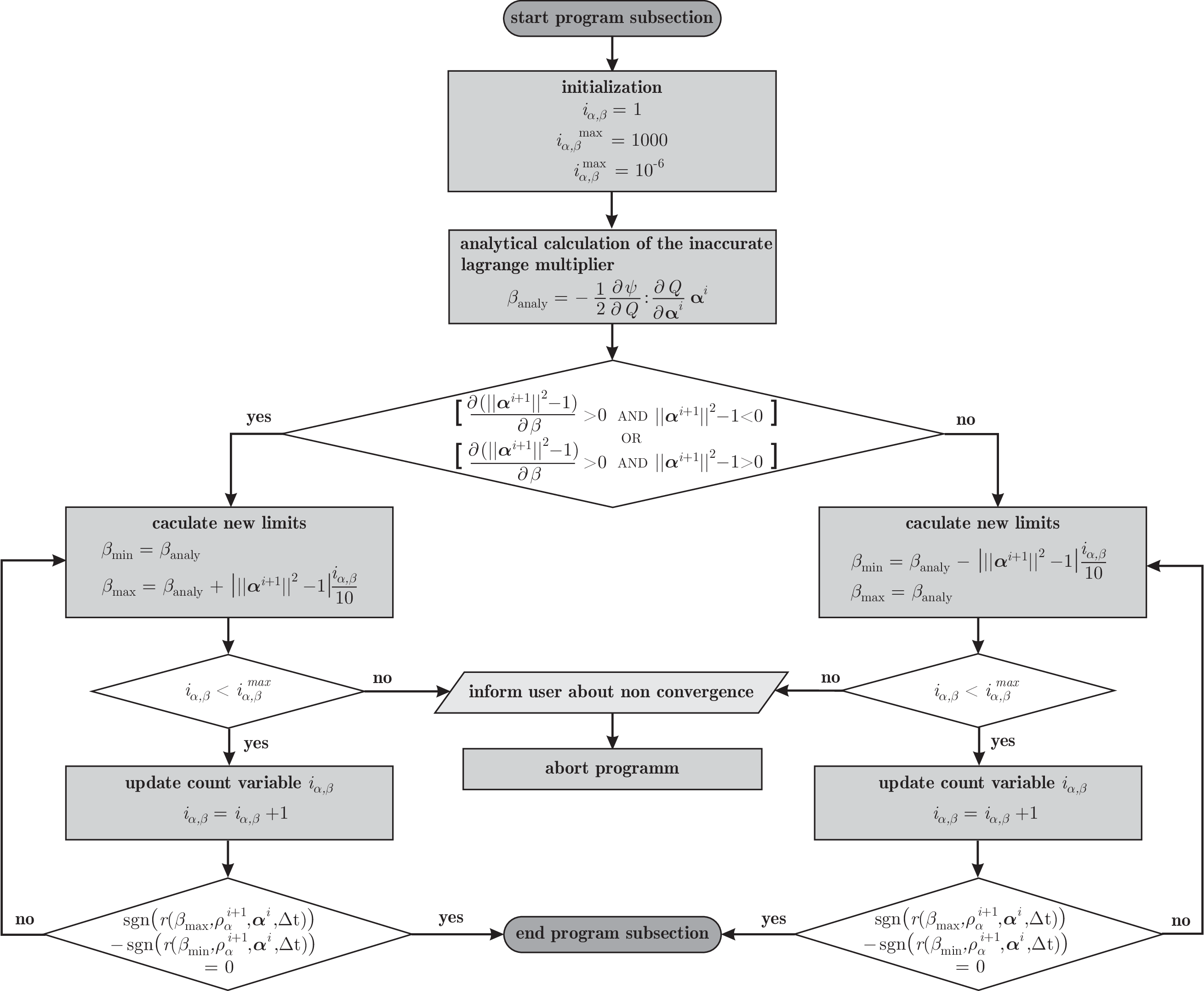}
\vspace{1mm}
\caption{Subsection of user subroutine UMAT to obtain limit values $\beta_{\mathrm{min}}$ and $\beta_{\mathrm{max}}$}\label{fig:ProgramSubsection_LimitValuesLagrange}
\end{figure}
\begin{figure}[H]%
\centering
\includegraphics[width=0.75\textwidth]{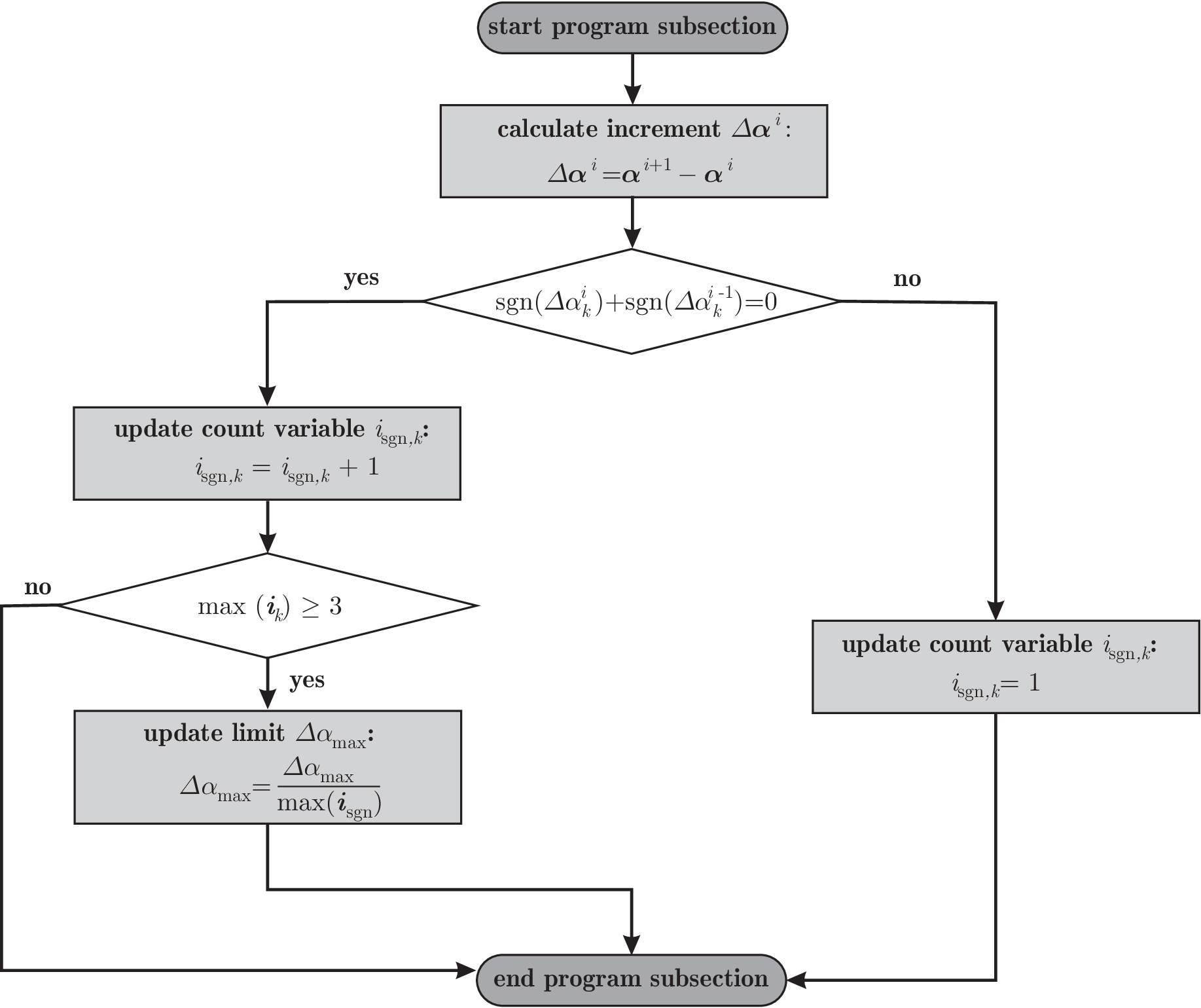}
\vspace{3mm}
\caption{Subsection of user subroutine UMAT to avoid oscillation when calculating Euler-Rodrigues paramter $\pmb{\alpha}^{n+1}$}\label{fig:ProgramSubsection_preventOscillation}
\end{figure}

\newpage
\section{Derivation of flow functions}\label{secA2}
\vspace*{5mm}

\textbf{For volume fractions $\boldsymbol{\lambda}$:}  \\
\vspace{-7mm}
\begin{align}
    \Delta^\ast_\lambda &= \underset{\dot{\lambda}\in \mathbb{R}}{\mathrm{sup}} \Bigg(\boldsymbol{\overline{p}}_\lambda \cdot \boldsymbol{\dot{\lambda}} - r_\lambda\ ||\boldsymbol{\dot{\lambda}}||\Bigg) \\  
    &= \underset{\dot{\lambda}\in \mathbb{R}}{\mathrm{sup}}
    \Bigg(\dfrac{||\boldsymbol{\dot{\lambda}}||}{r_\lambda} \Big[||\overline{\boldsymbol{p}}_\lambda||^2 -  r^2_\lambda\Big]\Bigg)\\
    &= \left\{ 
    \begin{array}{lll}
    \infty \quad & \text{if} \quad ||\overline{\boldsymbol{p}}_\lambda|| - r_\lambda   >  0 & \quad: ||\boldsymbol{\dot{\lambda}}|| \neq 0\\[2ex]
    0   \quad & \text{if} \quad ||\overline{\boldsymbol{p}}_\lambda|| - r_\lambda \leq 0 & \quad: ||\boldsymbol{\dot{\lambda}}||   =  0
    \end{array}
    \right.\\
    &\Rightarrow \quad \fbox{$\phi_{\lambda}:\, = ||\overline{\boldsymbol{p}}_\lambda|| - r_\lambda \leq 0$} 
\end{align}
\newline
\textbf{For Euler-Rodrigues parameter $\boldsymbol{\alpha}$:} \\
\vspace{-7mm}
\begin{align}
\Delta^\ast_\alpha&=\underset{\dot{\alpha}\in \mathbb{R}}{\mathrm{sup}} \Bigg(\boldsymbol{\overline{p}}_\alpha \cdot \boldsymbol{\dot{\alpha}} - r_\alpha\ ||\boldsymbol{\dot{\alpha}}||\Bigg)\\
&=  \underset{\dot{\alpha}\in \mathbb{R}}{\mathrm{sup}}
\Bigg(\dfrac{||\boldsymbol{\dot{\alpha}}||}{r_\alpha} \Big[||\overline{\boldsymbol{p}}_\alpha||^2 -  r^2_\alpha\Big]\Bigg)\\
&= \left\{ 
\begin{array}{lll}
\infty \quad & \text{if} \quad ||\overline{\boldsymbol{p}}_\alpha|| - r_\alpha   >  0 &  \quad: ||\boldsymbol{\dot{\alpha}}|| \neq 0\\
0   \quad & \text{if} \quad ||\overline{\boldsymbol{p}}_\alpha|| - r_\alpha \leq 0 &  \quad: ||\boldsymbol{\dot{\alpha}}||   =  0
\end{array}
\right.\\
&\Rightarrow \quad \fbox{$\phi_{\alpha}:\, = ||\overline{\boldsymbol{p}}_\alpha|| - r_\alpha \leq 0$}
\end{align}
\newline
\noindent \textbf{For plastic strains $\boldsymbol{\varepsilon}_{\mathrm{pl}}$:}\\
\vspace{-7mm}
\begin{align}
\Delta^\ast_\mathrm{pl}&= \underset{\dot{\varepsilon}_\mathrm{pl}\in \mathbb{R}}{\mathrm{sup}}  \Bigg(
\mathrm{dev}\,\boldsymbol{\sigma} : \boldsymbol{\dot{\varepsilon}}_\mathrm{pl} - 
\Big(\Delta 
(\boldsymbol{\dot{\varepsilon}}_\mathrm{pl}) + \mu \,c_{\alpha} \Big)
\Bigg)\\
& = \underset{\dot{\varepsilon}_\mathrm{pl}\in \mathbb{R}}{\mathrm{sup}}  
\Bigg(\dfrac{||\boldsymbol{\dot{\varepsilon}}_\mathrm{pl}||}{r_\mathrm{pl} + \mu} \Big[||\mathrm{dev}\,\boldsymbol{\sigma}||^2 -  (r_\mathrm{pl} + \mu)^2\Big]\Bigg) - \mu \, \dot{\kappa}
\\
&= \left\{ 
\begin{array}{lll}
\infty \quad \quad & \text{if}   \quad  ||\mathrm{dev}\,\boldsymbol{\sigma}|| - (r_\mathrm{pl} + \mu)   >  0 & : \quad ||\boldsymbol{\dot{\varepsilon}}_\mathrm{pl}|| \neq 0\\[3ex]
- \mu \, \dot{\kappa} \quad \quad  & \text{if} \quad   ||\mathrm{dev}\,\boldsymbol{\sigma}|| - (r_\mathrm{pl} + \mu)  \leq 0 & : \quad ||\boldsymbol{\dot{\varepsilon}}_\mathrm{pl}||   =  0
\end{array}
\right. \\
&\Rightarrow \quad \fbox{$\phi_{\mathrm{pl}}:\, = ||\mathrm{dev}\,\boldsymbol{\sigma}|| - (r_\mathrm{pl} + \mu) \leq 0
$}
\end{align}





\end{appendices}


\newpage
\bibliography{sn-bibliography}

\end{document}